\documentstyle[12pt,epsf]{article}

\epsfverbosetrue
\def\figlabel#1{\xdef#1{\thefigure}}
\def\fig#1{fig.~#1}

\def\figalign#1#2#3#4#5#6{
\begin{figure}
\centerline{
\hbox to 2.5truein{\vtop{\hsize=2.5truein\epsfxsize=6cm
\centerline{\epsfbox{#1} }
\caption[]{#3}
\figlabel{#2}
}}
\qquad\hbox to 2.5truein{\vtop{\hsize=2.5truein\epsfxsize=6cm
\centerline{\epsfbox{#4} }
\caption[]{#6}
\figlabel{#5}
}}
}
\end{figure}
}

\def\be{\begin{equation}}
\def\ee{\end{equation}}
\def\bea{\begin{eqnarray}}
\def\eea{\end{eqnarray}}
\def\G{\Gamma}

\def\a{\alpha}
\def\b{\beta}
\def\g{\gamma}
\def\d{\delta}
\def\pa{\partial}
\def\t{\widetilde}

\begin{document}

\begin{titlepage}

\begin{flushright}
{ ~}\vskip -1in
YCTP-P8-98\\
UB-ECM-PF 98/09\\
hep-th/9804038\\
March 1998\\
\end{flushright}

\vspace*{20pt}
\bigskip

\begin{center}
 {\Large DUALITY SYMMETRY IN \\
   \bigskip
\Large SOFTLY BROKEN ${\cal N}=2$ GAUGE THEORIES}
\vskip 0.9truecm
{\large Marcos Mari\~no$^{a}$ and Frederic Zamora$^{b}$.}

\vspace{1pc}

{\em $^a$ Department of Physics, Yale University\\
New Haven, CT 06520,
 USA.\\

\bigskip

$^b$ Departament d'Estructura i Constituents de la Materia,
\\ Facultat de F\'\i sica, Universitat de Barcelona,\\

Diagonal 647, E-08028 Barcelona, Spain.}\\

\vspace{5pc}

{\large \bf Abstract}

\end{center}

We study the soft breaking of ${\cal N}=2$ self-dual gauge theories
down to ${\cal N}=0$ by promoting the coupling constant and
hypermultiplet masses to spurions, and we analyze the microscopic
duality symmetry in the resulting models. Explicit formulae are given
for the Seiberg-Witten periods and couplings in the case of $SU(2)$,
and we perform a numerical study of the non-supersymmetric vacuum
structure in the case of the mass-deformed ${\cal N}=4$ $SU(2)$ gauge
theory. Although the softly broken model has a well-defined behavior
under the duality symmetry, the stable vacua are in a confining
phase. We also extend some of the results to the self-dual theories
with classical gauge groups, and we obtain the RG equation for these
models.

\end{titlepage}


\section{Introduction and Conclusions}

\setcounter{equation}{0}
\indent

After the exact results on the non-perturbative behavior of ${\cal
N}=1$ and ${\cal N}=2$ supersymmetric gauge theories in four
dimensions (see \cite{rev} for a review and extensive list of
references), it is natural to ask at which extent these results can be
extended to non-supersymmetric models. One possible avenue to explore
this issue is to consider the soft supersymmetry breaking of the
solved models with ${\cal N}=1$ \cite{softone} or ${\cal N}=2$
supersymmetry \cite{soft, amz, hsuangle} \footnote{An approach
suitable to explore the decoupling limit has been proposed in
\cite{francesco}.}.  The softly broken ${\cal N}=2$ models have been
studied in the case of asymptotically free theories and it has been
shown that they give interesting scenarios for confinement and chiral
symmetry breaking. They also have a very rich phase structure, and it
is possible to obtain exact pion Lagrangians up to two derivatives.

There is a very interesting class of ${\cal N}=2$ gauge theories that
has not been previously considered from the point of view of soft
supersymmetry breaking. These are the theories which have a zero beta
function at the perturbative level. We will call them self-dual gauge
theories, as they are supposed to have an exact duality symmetry
acting on the microscopic coupling constant, even when bare masses are
introduced for the matter multiplets (in this case, the duality
symmetry has also, in general, an action on the bare masses). This
symmetry in the space of microscopic theories, which relates theories
with different coupling constants, is the well-known Montonen-Olive
duality. In \cite{swtwo}, Seiberg and Witten showed that, for
self-dual theories with an $SU(2)$ gauge group, this microscopic
duality naturally appears in the effective description at low
energies, providing in this way a new check of the Montonen-Olive
conjecture.
 
In this paper we explore the behavior of the self-dual ${\cal N}=2$
gauge theories, once supersymmetry is broken down to ${\cal N}=0$ by
promoting the microscoping gauge coupling and the hypermultiplet
masses to spurions, and we study in detail the realization of the
duality symmetry in the softly broken theories. To understand some of
the new aspects arising in this class of models, it is useful to
recall the structure of the low-energy description of ${\cal N}=2$
gauge theories and of the softly broken models.
        
In the asymptotically free theories, duality appears only at the level
of the low-energy effective action. This means that at a particular
point in the moduli space we can describe the effective theory in {\it
any} duality frame, and the different frames will be related by ${\rm
Sp}(2r, {\bf Z})$ transformations, for a gauge group of rank $r$ (with
an inhomogeneous part in the case of the massive theories). In other
words, on the moduli space of vacua there is a flat ${\rm Sp}(2r, {\bf
Z})$ bundle and the ${\rm Sp}(2r, {\bf Z})$ group acts on the fibres
keeping the base points fixed. This will give different choices for
the sections of this bundle (locally represented by the $a_I(\vec u)$,
$a^I_D(\vec u)$ variables, $I=1, \dots, r$).  Of course, near the
singularities where BPS states become massless, there are preferred
duality frames in order to couple the hypermultiplets in a local way
to the corresponding vector multiplet. Therefore, the low-energy
effective actions of these theories are covariant under duality
transformations.

When the theory is softly broken with spurions, a crucial ingredient
for the consistency of the procedure is the compatibility with the
duality covariance of the effective theory.  This has been analyzed in
great detail in \cite{soft}, and derived in a general setting in
\cite{amz}.  The most important consequences of this analysis are the
following: first, the spurion superfields can be frozen in such a way
that their scalar components give the parameters already present in
the ${\cal N}=2$ theory, and the auxiliary components are
supersymmetry breaking parameters.  Second, the ${\cal N}=0$
`cosmological' term is invariant under duality transformations of the
effective theory. This means that at any point in the moduli space of
vacua, the value of this term does not depend on our choice of the
description with a duality frame.  As a consequence of this, the
effective potential (which equals the cosmological term plus the
contributions of the condensates) is globally defined on the moduli
space. 

 In the self-dual gauge theories, the prepotential depends now on the
microscopic gauge copling $\tau_0$ and on the masses of the
hypermultiplets $m_f$, $f=1, \cdots, N_f$. These are the parameters
that will be promoted to spurions, as in \cite{soft, amz}, in order to
softly break ${\cal N}=2$ supersymmetry down to ${\cal N}=0$. For a
fixed value of the microscopic coupling, the effective theory has
again a duality covariance which allows us to describe the low-energy
physics in different duality frames, and the effective potential must
be also invariant under these choices. As we will see, the above
mentioned requirements are again fulfilled, and the soft breaking of
the self-dual theories by promoting $\tau_0$, as well as the
microscopic masses, to spurions, is a consistent procedure.  But the
new ingredient appearing in the self-dual theories is the duality
symmetry acting on the microscopic coupling constant and on the bare
masses, as well as its interplay with the duality covariance in the
low-energy effective description. In the same way that the soft
breaking with spurions makes possible to preserve the information
encoded in the Seiberg-Witten solution, we will show that the behavior
under the microscopic duality symmetry is also under control in the
softly broken models. This is because the dependence on the
supersymmetry breaking parameters is encoded in the prepotential. In
particular, we will obtain the transformation under the duality
symmetry of all the generalized couplings arising in the softly broken
models, which are given by the second derivatives of the
prepotential. Thus we find non-supersymmetric models which preserve in
some way the duality invariance.

However, as we will show in the case of the mass-deformed ${\cal N}=4$
theory, the softly broken ${\cal N}=0$ model has a behavior rather
different from the behavior that one finds in the softly broken ${\cal
N}=1$ models analyzed in \cite{donwit}: there are generically runaway
vacua, except for a range of values of the microscopic coupling
constant in the strong coupling region $g^2_0 \sim 1$. In this range,
for a zero bare theta angle, there is only one single vacuum, {\it
i.e.} only one massive phase is realized for each value of the
microscopic coupling constant: the confining phase. For a non-zero
bare theta angle, the theory has again a single vacuum except for
$\theta_0= \pi$, where we have two degenerate vacua and a first order
phase transition, as in other softly broken models \cite{hsuangle,
konishi}. These results suggest that, although the microscopic duality
symmetry is under control in the softly broken ${\cal N}=2$ theory,
supersymmetry breaking triggers confinement (for small values of the
supersymmetry breaking parameter) and the duality symmetry of the
phase structure is lost.

The organization of the paper is as follows: In section 2, we review
the Seiberg-Witten solution for self-dual theories with the gauge
group $SU(2)$. Using the techniques introduced in \cite{amz}, we
present the explicit solution for the Seiberg-Witten differential and
periods.  In section 3 we analyze the softly broken model, and we
obtain the transformation properties of the couplings under the
microscopic duality symmetry. We also give explicit expressions for
all the couplings in the $SU(2)$ case. In section 4, we discuss the
softly broken ${\cal N}=2$ theories with classical gauge groups, and
we compute the dual spurion in these models.  This gives the RG
equation for the self-dual theories. In section 5, we discuss the
structure of the mass deformed ${\cal N}=4$ Yang-Mills theory.
Finally, in section 6, we study in detail the vacuum structure of the
softly broken, mass deformed ${\cal N}=4$ Yang-Mills theory.


\bigskip

\section{The Seiberg-Witten Solution for Self-Dual 
${\cal N}=2$, $SU(2)$ Gauge Theories}
\setcounter{equation}{0}
\indent

In this section we briefly review some of the properties of the
self-dual ${\cal N}=2$ theories with gauge group $SU(2)$ considered by
Seiberg and Witten in \cite{swtwo}. We also present explicit solutions
for the Seiberg-Witten differential $\lambda_{SW}$ and for the periods
$a$, $a_D$.

\subsection{The massless case.}
\indent

${\cal N}=2$ QCD with four massless flavours in the fundamental
representation, and ${\cal N}=4$ super Yang-Mills theories, have a
zero beta function at the perturbative level. When the vacuum is
conformally invariant, the theory is believed to be in a non-abelian
fixed point. Few things are known about this interacting
superconformal field theory\footnote{Recently, an interesting  
conjecture \cite{maldacena} 
allows to obtain results for the ${\cal N}=4$ superconformal 
Yang-Mills theory in the large $N_c$ limit.}. However, the theory is  
well understood
when the scalar component of the ${\cal N}=2$ vector multiplet gets a
vacuum expectation value, $a\not=0$ and conformal invariance is
spontaneously broken.  At scales lower than $|a|$, the theory is
described by an ${\cal N}=2$ $U(1)$ gauge theory that flows to a
trivial infra-red fixed point.  In fact there is an exact marginal
direction, of complex dimension one, for these fixed points. It is
parametrized by the complex number
\bea \tau_{0} &=& {\theta \over
\pi} + {8 \pi i \over g^2}, \,\,\,\,\,\ N_f=4, \nonumber\\ &=&  
{\theta
\over 2 \pi} + {4 \pi i \over g^2}, \,\,\,\,\,\ {\cal N}=4.
\label{coupling}
\eea
which is the value of the holomorphic $U(1)$ gauge coupling in the 
infra-red limit. Furthermore, the theory has an infinite number of  
stable BPS states
characterized by the relatively prime integers $(n_m, n_e)$, which  
are
related with the magnetic and electric charges of the $U(1)$ gauge
symmetry. Their mass spectrum is 
\be
{\cal M}(n_m, n_e) = \sqrt{2}|n_m\tau_0 + n_e||a|
\ee
In \cite{swtwo} it was assumed that, in the massless theories, the  
effective $U(1)$ coupling 
constant $\tau_{0}$ equals the bare coupling constant of the 
non-abelian microscopic theories, and that the classical formulae 
\bea 
a&=&{1 \over 2} {\sqrt 2u}, \,\,\,\,\,\ N_f=4, \nonumber\\
&=& { \sqrt 2u}, \,\,\,\,\,\  {\cal N}=4, \nonumber\\
a_D&=& \tau_{0} a,
\label{semiperiods}
\eea
where $u = \langle {\rm Tr} \phi^2 \rangle$, were not corrected  
quantum-mechanically. The prepotential can be written then 
as 
\be
{\cal F} (a, \tau_0)= {1 \over 2} \tau_0 a^2
\label{semiprep}
\ee
in both cases. 

The classical answers (\ref{semiperiods}) can be encoded in terms of  
an 
elliptic curve and a holomorphic one-form on it. Consider a lattice  
$\Gamma \subset 
{\bf C}$ generated by $\omega_1 = \pi$, $\omega_2 = \pi \tau_0$.  
Using  
the Abel map 
we can describe the torus ${\bf C}/\Gamma$ by an elliptic curve  
written in 
the Weierstrass form
\be
y_0^2=4x_0^3-g_2x_0-g_3=4(x_0- e_1)(x_0- e_2)(x_0- e_3),
\label{we}
\ee
where the roots $e_i$, $i=1,2,3$, are given in terms of theta  
functions \footnote{
Notice that we follow the conventions of \cite{mw} for the  
definitions of the 
theta functions.} by:
\bea
e_{1}&=&{1 \over 3}(\vartheta_4^4 (\tau_0) + \vartheta_3^4(\tau_0)),  
\nonumber\\
e_{2}&=&-{1 \over 3}(\vartheta_2^4 (\tau_0) + \vartheta_3^4(\tau_0)),  
\nonumber\\
e_{3}&=&{1 \over 3}(\vartheta_2^4 (\tau_0) - \vartheta_4^4(\tau_0)). 
\label{es}
\eea

Introducing the variables $x=ux_0$, $y={1 \over 2} u^{3/2}y_0$, we  
can write 
the elliptic curve associated to the massless $N_f=4$ or ${\cal N}=4$  
theories as 
\be
y^2=(x-e_1 (\tau_0) u) (x-e_2 (\tau_0) u) (x-e_3 (\tau_0) u),
\label{semicurve}
\ee
with the abelian differential
\bea
\omega&=&{ {\sqrt {2 /u}}  \over 4 \pi} {dx_0 \over y_0}= {{\sqrt 2}  
\over 8 \pi}{ dx \over y}, \,\,\,\,\,\ N_f=4, \nonumber\\
\omega&=&{ {\sqrt {2 /u}}  \over 2 \pi} {dx_0 \over y_0}= {{\sqrt 2}  
\over 4 \pi}{ dx \over y}, 
\,\,\,\,\,\  {\cal N}=4.
\label{abdiff}
\eea 
One can easily check that the periods of this differential are  
precisely $da_D/du$ and $da/du$.
In this way, the modulus of the torus described by (\ref{semicurve})
will be the $U(1)$ low energy massless coupling $\tau_0$.
If we consider the above curve with coupling $-1/\tau_0$, we obtain  
the same elliptic
curve after rescaling  $x_0 \rightarrow \tau_0^2 x_0$
and $y_0 \rightarrow \tau_0^3 y_0$, since
\bea
e_{1} \left(-{1 \over \tau_0}\right) &=& \tau_0^2 e_{2}(\tau_0),
\\
e_{3} \left(-{1 \over \tau_0} \right) &=& \tau_0^2 e_{3}(\tau_0).
\eea
Therefore, this is a symmetry of the vacuum. 
As a symmetry, it leaves invariant the spectrum,
but in a non-trivial way, since it maps the BPS state $(n_m, n_e)$  
for
coupling $\tau_0$ and expectation value $a$
to the BPS state $(n_e, -n_m)$ for coupling $-1/\tau_0$ and  
expectation 
value $\tau_0 \, a$. It is called an $S$-duality symmetry.
There is still another symmetry in the $\tau_0$ plane, called  
$T$-duality:
\bea
e_{1}(\tau_0 + 1) &=& e_{1}(\tau_0),
\\
e_{ 2}(\tau_0 \pm 1) &=& e_{ 3}(\tau_0).
\eea
It maps the state $(n_m, n_e)$ to the state $(n_m, n_e - n_m)$ with 
the same VEV $a$.

\subsection{The Seiberg-Witten curve.}
\indent

We will now briefly review the solution proposed in \cite{swtwo} for
the low-energy desciption of the conformally invariant theories
$N_f=4$ and ${\cal N}=4$, once scale invariance is broken explicitly
by the inclusion of bare masses for the hypermultiplets. Recall that
the ${\cal N}=4$ theory can be considered as an ${\cal N}=2$ super
Yang-Mills theory with a matter hypermultiplet in the adjoint
representation. When this hypermutiplet is given a mass $m$, the
resulting theory (which we will call, following current terminology,
mass-deformed ${\cal N}=4$ theory) has only ${\cal N}=2$
supersymmetry. In this case there is a moduli space of inequivalent
vacua and the mass formula for BPS states is a complicated function of
the dimensionless quantity $a/m$.

We first focus on the $N_f=4$ theory, where the duality invariance  
involves a group action on the bare masses. The massless theory has a  
$SO(8)$ flavour symmetry 
and we can consider that ${\vec m}^t=(m_1, \cdots, m_4)$ is in the  
adjoint representation 
of ${\rm Spin}(8)$. The ${\bf S}_3$ automorphism of ${\rm Spin}(8)$  
is generated by
two transformations that we will call $t^T$ and $t^S$.  They act on  
the masses as
\be
m_f \rightarrow t_{fg}^{T,S} m_g, 
\ee
where 
\be
t_{fg}^{T}=\left(\begin{array}{cccc}  1 &  0& 0& 0  \\ 0 &1& 0& 0\\  
0& 0& 1&0 \\ 
0&0&0&-1\end{array}\right), \,\,\,\,\ 
t_{fg}^{S}={1 \over 2} \left(\begin{array}{cccc}  1 &  1& 1& 1  \\ 1  
&1& -1& -1\\ 1& -1& 1&-1 \\ 
1&-1&-1&1\end{array}\right).
\ee
Notice that the matrices in ${\bf S}_3$ are symmetric. 

We now introduce the variable 
\be
{\tilde u}=u-{1 \over 2 }e_1 R 
\label{uvar}
\ee
as a parameter on the moduli space of vacua, where $R$ is given by
\be
R={1\over 2} \sum_f m_f^2.
\label{r}
\ee
 In terms of this  
variable, the Seiberg-Witten curve for $N_f=4$ can be written as
\be
y^2= W_1W_2W_3
+ A\left(W_1T_1(e_2-e_3)+W_2T_2(e_3-e_1)+W_3T_3(e_1-e_2) \right)
- A^2 N,
\label{curva}
\ee
where
\bea
W_i &=&x-e_i \tilde u - e_i^2 R, \nonumber \\
A&=&(e_1-e_2)(e_2-e_3)(e_3-e_1), \nonumber\\
T_1&=&{1 \over 12}\sum _{f>g}m_f^2m_g^2 - {1\over 24}\sum_f m_f^4,   
\nonumber\\
T_2&=&-{1 \over 2}\prod_f m_f - {1 \over 24}\sum _{f>g}m_f^2m_g^2 +
{1\over 48}\sum_f m_f^4, \nonumber\\
T_3&=&{1 \over 2}\prod_f m_f - {1 \over 24}\sum _{f>g}m_f^2m_g^2 +
{1\over 48}\sum_f m_f^4, \nonumber\\
N&=&{3 \over 16}\sum_{f>g>h}m_f^2m_g^2m_h^2 -
{1 \over 96}\sum_{f \not= g} m_f^2m_g^4 + {1 \over 96}\sum_f m_f^6. 
\label{parametros}
\eea
$R$ and $N$ are invariant under ${\bf S}_3$, while the $T_i$,  
$i=1,2,3$, are permuted in 
the following way:
\be
t^T: \begin{array}{ccc} T_1 & \leftrightarrow & T_1 \\
  T_2  & \leftrightarrow &T_3 \end{array}, \,\,\,\,\,\,\,\ 
t^S: \begin{array}{ccc} T_1 & \leftrightarrow & T_2 \\
  T_3  & \leftrightarrow &T_3 \end{array}.
\label{sthree}
\ee 

The curve (\ref{curva}) has an invariance group given by the
semidirect group $SL(2, {\bf Z})\times {\bf S}_3$, generated by the
two elements $(T, t^T)$ and $(S, t^S)$, if one also assumes that $y$,
$x$ and $\tilde u$ are modular forms of $SL(2, {\bf Z})$ of modular
weights $6$, $4$ and $2$, respectively.  Therefore, $SL(2, {\bf Z})$
duality is mixed with $SO(8)$ triality acting on the masses.  Notice
that (\ref{uvar}) relates the modulus ${\tilde u}$ of the
Seiberg-Witten curve to the gauge order parameter $u= \langle {\rm Tr}
\phi^2 \rangle$. It has been observed in \cite{dorey} that for $N_f=4$
the coupling $\tau_{0}$ appearing in the Seiberg-Witten solution
differs from the microscopic coupling, due to instanton corrections
(for the mass-deformed ${\cal N}=4$ theory \cite{doreydos}, the
effective gauge coupling does not receive any instanton correction due
to ${\cal N}=4$ supersymmetry). Also \cite {dorey, doreydos}, the
label $u$ for the moduli space is related to $\langle {\rm Tr}\phi^2
\rangle$ by an expression which differs from the one given in
\cite{swtwo}, both in the $N_f=4$ and in the mass-deformed ${\cal
N}=4$ cases. We then have a good label on the moduli space of the
curve, which is in fact a modular form, but due to instanton
corrections (which are not predicted by the Seiberg-Witten curve) we
do not know how to relate it to the order parameter of the underlying
microscopic theory. We will again follow the notation in \cite{swtwo}
and keep the relation (\ref{uvar}), although as we have pointed out,
the $u$ variable defined in this way is not the original order
parameter.
 
The curve (\ref{curva}) can be written in a very convenient way if we  
redefine the variable $x$ as follows
\be
x \rightarrow x + e_1 u + {1 \over 2} e_1^2 R,
\label{redef}
\ee
We then obtain the following expression for the Seiberg-Witten curve:
\be
y^2=x(x-\alpha u)(x-\beta u) +ax^2+bx+cu +d,
\label{otracurva}
\ee
where
\bea
a& =&-{1\over 4} (\alpha-\beta)^2  \sum_f m_f ^2,\nonumber\\
                    b& =& -{1 \over 4}(\alpha-\beta)^2  
\alpha\beta\sum_{f<g}
                          m_f^2m_g^2 +{1 \over 2}\alpha\beta
 (\alpha^2-\beta^2)\prod_{f=1}^4m_f,\nonumber\\
                    c& =& -
           (\alpha-\beta)\alpha^2\beta^2\prod_{f=1}^4m_f,
\nonumber\\                    
d& =& -{1 \over 4} (\alpha-\beta)^2\alpha^2\beta^2
                    \sum_{f<g<h}m_f^2m_g^2m_h ^2,
\label{coefcs}
\eea
and 
\be
\alpha = e_2- e_1= -\vartheta_3^4 (\tau_0), \,\,\,\,\,\, \beta=e_3-  
e_1=-\vartheta_4^4(\tau_0).
\label{albe}
\ee 
This form is very useful to find the explicit expression for the
abelian differential $\lambda_{SW}$, as only the power $u^2$ appears
in it. Actually we will find more convenient to work with the shifted
variable $\tilde u$, as this variable is a true modular
form. Therefore we will mainly consider the curve (\ref{otracurva})
but expressed in terms of ${\tilde u}$.  The curve for the
mass-deformed ${\cal N}=4$ theory is a special case of (\ref{curva})
and (\ref{otracurva}) for the values of the masses $m_1=m_2 =m/2$,
$m_3=m_4=0$. One has also to take into account the different
normalization for the abelian differential.

\subsection{The Seiberg-Witten Differential and Periods.}
\indent

In the $N_f=4$ case, the Seiberg-Witten abelian differential  
$\lambda_{SW}$ is determined, as usual,  
by the equation
\be
{\partial  \lambda_{SW} \over \partial \tilde u}={ {\sqrt{2}} \over 8  
\pi} 
{dx \over y},
\label{swabel}
\ee
as this requirement guarantees the positivity of the metric.   
The computation of this differential can be done as in the $N_f=3$ 
case in \cite{amz}. First of all we write the curve 
(\ref{otracurva}) as 
\be
y^2= p^2 {\tilde u}^2 + B \tilde u + C, 
\label{ucurve}
\ee
where
\bea
p^2 &=& \alpha \beta x, \nonumber\\
B&=& c- (\alpha +\beta) x^2 + p^2 e_1 R, \nonumber\\
C&=& x^3 + a x^2 + b x + d + {1 \over 2} e_1 R B + {1 \over 4} p^2  
e_1^2 R^2.
\label{coefabel}
\eea
Integrating w.r.t. $\tilde u$, and up to an exact differential, we  
find 
the expression 
\be
\lambda_{SW} = -{ {\sqrt 2} \over 8 \pi} {dx \over y}  \biggl( 
2 \tilde u  + { x B' C - {1 \over 2} (x C' B + BC) -\tilde u x (p^2  
C- {1 \over 4} B^2)' 
\over p^2 C- {1 \over 4} B^2} \biggr).
\label{explicit}
\ee
The expression in the denominator is the discriminant of the curve  
(\ref{ucurve})
as a polynomial in $\tilde u$, and is given by
\be
p^2 C- {1 \over 4} B^2 = -{1 \over 4} (\alpha-\beta)^2 \prod_{f=1}^4  
(x + \alpha \beta m_f^2).
\label{discr}
\ee
The position of the poles are then $x_f= -\alpha \beta m_f^2$, as it  
has been 
derived in section 17 of \cite{swtwo}. After some straightforward  
computations,
we obtain
\be
\lambda_{SW}= - { {\sqrt 2} \over 8 \pi}{dx \over y} \biggl( -2 (  
\tilde u - e_1 R) + \alpha \beta 
\sum_{f=1}^4 { { \alpha + \beta \over 2} m_f^4 + m_f^2 (\tilde u + {1  
\over 2} e_1 R) 
+ {\alpha- \beta \over 2} \prod_{f} m_f  \over x + \alpha \beta  
m_f^2} 
\biggr).
\label{masex}
\ee
The numerators in the poles have the form $q_f m_f$, where 
$q_f$ verifies 
\be
y^2 (-\alpha \beta m_f^2) = -q_f^2.
\label{numer}
\ee

We can then compute the matrix $S_{n}^f$ associated 
to the massive Seiberg-Witten theories 
with $N_f$ flavours and $N_p$ simple poles in $\lambda_{SW}$, which  
is 
defined  through the relation \cite{amz}
\be
{\rm Res}_{x=x_n} \lambda_{SW} = {1 \over 2\pi i} \sum_{f=1}^{N_f}  
S_n^f { m_f \over {\sqrt 2}}.
\label{residues}
\ee
We see from (\ref{masex}) that, in the $N_f=4$ theory, $S_n^f=  
-{1\over 2} \delta_n^f$. 
   
Once the Seiberg-Witten differential is known, we can compute its  
periods, 
$a$ and $a_D$, using uniformization as in \cite{amz} (see also 
\cite{ferrari, bf}). We then write the curve (\ref{curva}) in  
Weierstrass form,  
\be
Y^2=4X^3-g_2X-g_3=4(X-\hat e_1)(X-\hat e_2)(X-\hat e_3). 
\label{weier}
\ee
The original Seiberg-Witten curves (\ref{curva}), (\ref{otracurva})  
have the structure 
\be
y^2=x^3+  a_2 x^2+a_1x +a_0 ,
\label{curvasw}
\ee
where the coefficients $a_i$, $i=0,1,2$, depend on $\tilde u$, on the  
bare 
coupling constant $\tau_0$, and on the bare 
quark masses $m_f$, $f=1, \cdots, 4$. To put this curve in the  
Weierstrass form, 
it suffices to redefine the variables as 
\be
y=4Y, \,\,\,\,\,\,\,\ x=4X-{1 \over 3}a_2,
\label{cambio}
\ee
where $a_2$ is given by:
\be
a_2= -(\alpha + \beta) {\tilde u} + \alpha \beta R -{1 \over 3} R E_4  
(\tau_0),
\label{atwo}
\ee
and $E_4(\tau_0)$ is the Eisenstein series, which can be expressed 
in terms of theta functions as
\be
E_4(\tau_0)= {1 \over 2}(e_1^2 + e_2^2 + e_3^2)= {1\over 2} (  
\vartheta_2^8 + 
\vartheta_3^8 + \vartheta_4^8).
\label{efour}
\ee

Notice that the roots $\hat e_i$ in (\ref{weier}) are deformations of  
the roots $(1/4)e_iu$ of the 
curve (\ref{semicurve}) (after writing this curve in Weierstrass  
form). These
deformations are parametrized by the bare masses of the  
hypermultiplets. The Seiberg-Witten periods can be computed in terms  
of the periods of $dX/Y$, which we take 
as  
\bea
 \omega_1 &= & \oint _{\alpha_1}{ dX \over Y} = \int_{\hat e_2}^{\hat  
e_3} {dX \over 
\sqrt{(X-\hat e_1)(X-\hat e_2)(X-\hat e_3)} },\nonumber \\ 
\omega_2 &= & \oint _{\alpha_2}{ dX \over Y}=\int_{\hat e_1}^{\hat  
e_3} {dX \over 
\sqrt{(X-\hat e_1)(X-\hat e_2)(X-\hat e_3)}}.
\label{periodos}
\eea

We integrate now the Seiberg-Witten differential on these cycles. To  
do this, we 
have to introduce the variables $z_f$, which correspond to the poles  
$x_f$ under
the Abel map. They are defined by:
\be
4 \wp (z_f) + {\alpha + \beta \over 3} \tilde u + {1\over 9} R E_4  
(\tau_0) + \alpha \beta \bigl( 
m_f^2 - {1 \over 3}R \bigr) =0.
\label{zpolos}
\ee
The periods of the Seiberg-Witten diferential for $N_f=4$ are then  
given by
\be
a_i = {{\sqrt 2} \over 4 \pi} \biggl( (\tilde u - e_1 R) \omega_i + i  
\sum_{f=1}^4 
m_f [ \omega _i \zeta (z_f) - 2 z_f \zeta ({\omega_i \over 2})]  
\biggr), \,\,\,\,\,\,\ i =1,2.
\label{swp} 
\ee
In the case of zero masses, we see that $\omega_1 = 2 \pi u^{-1/2}$, 
$\omega_2= 2 \pi \tau_0  u^{-1/2}$, 
and we recover (\ref{semiperiods}). Also, in the case of  
the mass-deformed ${\cal N}=4$ theory, 
the expression (\ref{swp}) agrees with that given in \cite{ferrari}.


\bigskip

\section{Soft Breaking of the Self-Dual Theories}
\setcounter{equation}{0}
\indent

The ${\cal N}=2$ prepotential ${\cal F}(a, \tau_0, m_f)$ of the 
$N_f=4$ theory can be obtained from the curve (\ref{curva}), and 
the extension of the VEV $a$ to 
an ${\cal N}=2$ $U(1)$ vector superfield allows
to study the low energy dynamics of the theory
as a function of the microscopic parameters $\tau_0$, $m_f$
and the modulus $\tilde u$ (which gives the VEV $a$).
The ${\cal N}=2$ soft breaking technique 
consists in promoting the microscopic parameters $\tau_0$
and $m_f$ to the status of ${\cal N}=2$ frozen vector superfields
(spurions), with non-zero auxiliary fields:
\bea
\tau_0 \quad \rightarrow && \left( \ {\cal T}_0 = \tau_0 + \theta^2  
F_0\ , 
\ \ V_0 = {1 \over 2} D_0 \theta^2 {\bar \theta}^2 \ \right), 
\nonumber
\\
m_f \quad \rightarrow && \left(
M_f= {m_f \over {\sqrt 2}}+ \theta^2 F^f\ , 
\ \ V_f= {1 \over 2} D^f \theta^2 {\bar \theta}^2 \right)
\label{vevs}
\eea
In this way, we break supersymmetry down to ${\cal N}=0$, but we  
still keep all the 
nonperturbative information of the Seiberg-Witten curve.
The flat direction parametrized by $\tilde u$
is lifted and there is usually a unique 
non-supersymmetric physical vacuum. 
In \cite{soft, amz}, it was generally located 
near a Seiberg-Witten singularity, forcing the associated
light hypermultiplet(s) to condense.

\subsection{The effective potential.}
\indent

Near the Seiberg-Witten singularities 
where $k$ mutually local hypermultiplets
$(H_i, {\t H}_i)$, $i=1, ...,k$ become massless, 
with $ a + S_i^f m_f/\sqrt{2} =0$, we have the effective potential  
\cite{amz}
\bea
V_{\rm eff}&=&\Bigl({b_{aA} b_{aB} \over b_{aa}}-b_{AB}\Bigr) 
\Bigl({1 \over 2}D_A D_B + F_A {\overline F}^B \Bigr) + 
{b_{aA} \over b_{aa}}D_A \sum_{i=1}^{k} (|h_i|^2-|{\widetilde  
h}_i|^2) 
\nonumber \\ 
&+& { \sqrt{2} b_{aA}\over b_{aa}} \Bigl( F_A
  h_i{\widetilde h}_i + {\overline F}^A 
{\overline h}_i{\overline {\widetilde h}}_i \Bigr) + 
{2 \over b_{aa}} {\overline h}_i{\overline {\widetilde h}}_i
h_j{\widetilde h}_j \nonumber\\
&+& {1 \over 2 b_{aa}}\sum_{i,j=1}^{k} (|h_i|^2-|{\widetilde h}_i|^2)
(|h_j|^2-|{\widetilde h}_j|^2) -  
 D_f S^f_i (|h_i|^2-|{\widetilde h}_i|^2) \nonumber\\
&+& 2|a + S^f_i {m_f \over {\sqrt 2}}|^2(|h_i|^2+|{\widetilde  
h}_i|^2) 
\nonumber 
\\
&-&
\sqrt{2} \Bigl( S^f_i F_f h_i{\widetilde h}_i+ 
S^f_i{\overline F}^f {\overline h}_i{\overline {\widetilde h}}_i   
\Bigr),
\label{twovi}
\eea
where $A, B=0, 1, ..., f$ correspond to the spurion indices
and $b_{IJ}=(1/4\pi) {\rm Im(}\tau^{IJ})$ come from the matrix of  
holomorphic couplings
\footnote{Some of these couplings have a  
nice physical interpretation. In 
\cite{kon} it has been shown that Re($\tau^{af})$ 
is related to the physical  
baryon numbers of the BPS states by a generalized Witten effect. 
On the other hand, in the topological version of the ${\cal N}=2$ 
theory, the coupling $\tau^{00}$ appears as a contact term for  
a family of operators \cite{mw, lns, mm}.}
\bea
\nonumber
\tau^{aa}= \frac{\partial^2 {\cal F}}{\partial a^2}, \quad
\tau^{fa}= {\sqrt 2}\frac{\partial^2 {\cal F}}{\partial a \partial  
m_f}, \quad
\tau^{0a}= \frac{\partial^2 {\cal F}}{\partial a \partial \tau_0},  
\quad
\\
\tau^{fg}= 2\frac{\partial^2 {\cal F}}{\pa m_f \pa m_g}, \quad
\tau^{0f}= {\sqrt 2}\frac{\partial^2 {\cal F}}{\pa m_f \pa \tau_0},  
\quad
\tau^{00}= \frac{\partial^2 {\cal F}}{\pa \tau_0^2}. 
\label{coupl}
\eea

\subsection{Behavior under the duality symmetry.}
\indent

The key ingredient to analyze the behavior of the soflty broken model 
is the fact that a microscopic $SL(2,{\bf Z})$ transformation  
(combined with triality) 
induces an inhomogeneous $SL(2, {\bf Z})$ transformation in the  
effective theory, which 
in general will be different from the microscopic one, because of  
possible 
monodromy transformations \cite{ferrari}. In the microscopic 
theory, the $SL(2, {\bf Z})\times {\bf S}_3$ acts on the coupling  
and masses as 
\bea
\tau_0 \rightarrow \tau_0^{\Gamma} &=& {\alpha \tau_0 + \beta \over  
\gamma \tau_0 + \delta},
\nonumber\\
m_f \rightarrow m_f^\Gamma &=& t_{fg}m_g.
\label{micros}
\eea 
 This will induce the following effective duality transformation,
\be
\left(\begin{array}{c} a_D \\ a \end{array}\right) \rightarrow 
\left(\begin{array}{c} a^{\Gamma}_D \\ a^{\Gamma} \end{array}\right)  
= \left(\begin{array}{cc}  \hat \alpha &  \hat \beta  \\ \hat \gamma  
&  
\hat \delta \end{array}\right) \left(\begin{array}{c} a_D \\ a  
\end{array}\right) + \left(\begin{array}{c} p^fm_f/{\sqrt 2} \\  
q^fm_f /{\sqrt 2}\end{array}\right),
\label{macros}
\ee
where we understand 
$a^\Gamma_i = a_i ((\g \tau_0 + \d)^2 {\t u}, \tau_0^\Gamma,  
m^\Gamma)$. The variables $a$, $a_D$ correspond to an 
arbitrary frame with a local prepotential ${\cal F}(a, \tau_0, m_f)$  
and 
$a_D=\partial {\cal F} /\partial a$.   

We choose as our independent variables for the prepotential $a$,  
$\tau_0$ and $m_f$. To analyze the softly broken model, we introduce  
the generalized dual variables (that we will call dual spurions) as
\bea
\tau_{0,D}&=& {\partial {\cal F} \over \partial \tau_0},\nonumber\\
m^f_D&=& {\sqrt 2}  {\partial {\cal F} \over \partial m_f}.
\label{spduals}
\eea
Using (\ref{micros}) and (\ref{macros}) we find the relations 
\bea
\Big( {\partial \over \partial a} \Big)_{\G}
&=&{1 \over \hat \gamma \tau^{aa} +\hat \delta }{\partial \over  
\partial a},
\label{deriva}
\\
\Big( {\partial \over \partial \tau_0} \Big)_{\G}
&=&(\gamma \tau_0 + \delta)^2 \Bigl({\partial \over \partial  
\tau_0}-{\hat \gamma \tau^{a0} \over  \hat \gamma
\tau^{aa} +\hat \delta}{\partial \over \partial a}\Bigr),
\label{derivs}
\\
\Big( {\partial \over \partial m_f} \Big)_{\G}
&=&t_{gf}^{-1} \Bigl({\partial \over \partial m_g}-{1 \over {\sqrt  
2}} 
\frac{q^g+\hat \g \tau^{ag}}{\hat \g \tau^{aa} +\hat \d}
{\partial \over \partial a}\Bigr).
\label{derivm}
\eea
On the other hand, the transformation of the prepotential under  
(\ref{macros}) is the 
usual one,  
\bea
{\cal F}^{\Gamma}(a^{\Gamma}, m_f) &=& {\cal F}(a, m_f) + {1 \over 2}  
\hat \beta 
\hat \delta a^2 + {1 \over 2} \hat \alpha \hat \gamma a^2_D + \hat  
\beta \hat \gamma  a a_D  
\nonumber
\\
\label{genFt}
&+&  p^f {m_f \over {\sqrt 2}}(\hat \gamma a_D + \hat \delta a).
\eea
Acting with  (\ref{derivs}) and (\ref{derivm}) on ${\cal F}^{\Gamma}$
 one obtains the transformations of the dual 
spurions $m^f_D$ and $\tau_{0,D}$:
\bea
 m^{\Gamma ,f} _D &=& \Big( {\partial {\cal F}^{\Gamma} \over  
\partial m_f}
 \Big)_{\G} \nonumber 
\\
&=& t_{gf}^{-1} \bigl( m^g_D + p^g(\hat \g a_D + \hat \d a) -  
q^g(\hat \a a_D +\hat  \b a) - q^g p^h {m_h \over {\sqrt 2}}\bigr),
\nonumber\\
\tau^{\G}_{0,D}&=& \Big( {\partial  {\cal F}^{\Gamma} \over \partial  
\tau_{0}}
 \Big)_{\G} = (\gamma \tau_0 + \delta)^2 \tau_{0,D}.
\label{duals}
\eea
We then see that the dual spurion transforms as a modular form of  
weight 
$(2,0)$ under the microscopic duality. 

We can also obtain the transformations for all the 
low energy couplings
(\ref{coupl}) under a microscopic duality transformation, using  
($3.7-9$): 
\bea
( \tau^{\Gamma})^{aa} &=& \frac{\hat \a \tau^{aa} + \hat \b}{\hat \g  
\tau^{aa} +\hat  \d}, \quad
\nonumber\\
( \tau^{\Gamma})^{0a} &=& \frac{(\g \tau_0 + \d)^2 }{\hat \g  
\tau^{aa} + \hat \d}\tau^{0a}, \quad
\nonumber\\
( \tau^{\Gamma})^{00}&=&(\g \tau_0 + \d)^4 \biggl( \tau^{00} -  
\frac{\hat \g (\tau^{0a})^2}{\hat \g \tau^{aa} + \hat \d} \biggr) + 2  
\g (\g \tau_0 + \d)^3 \tau_{0,D},
\nonumber\\
( \tau^{\Gamma})^{af} &=&t_{gf}^{-1}\biggl( \frac{\tau^{ag}}{\hat \g  
\tau^{aa} + \hat \d} -q^g (\frac{\hat \a \tau^{aa} 
+\hat  \b}
{\hat \g \tau^{aa} + \hat \d}) +p^g \biggr), \quad
\nonumber\\
( \tau^{\Gamma})^{f0} &=&(\g \tau_0 + \d)^2 t_{gf}^{-1} \biggl(   
\tau^{g0} - (\frac{q^g+\hat \g \tau^{ga}}{\hat \g \tau^{aa} +\hat  
\d})
 \tau^{0a}\biggr) ,
\nonumber\\
( \tau^{\Gamma})^{fg} &=&t_{hf}^{-1} t_{kg}^{-1} \biggl(  \tau^{kh} -  
\frac{\hat \g \tau^{ka} \tau^{ha}}{\hat \g \tau^{aa} +\hat  \d}
 - \frac{q^k\tau^{ha}}{\hat \g \tau^{aa} + \hat \d} - \frac{q^h  
\tau^{ka}}{\hat \g \tau^{aa} + \hat \d}
\nonumber\\
 & &- p^hq^k - p^k q^h + q^k q^h (\frac{\hat \a \tau^{aa} 
+ \hat \b}{\hat \g \tau^{aa} + \hat \d})\biggr) .
\label{monotrans}
\eea

As we said in the introduction, it was shown in \cite{amz} that the
${\cal N}=2$ soft breaking preserves the macroscopic duality
covariance of the theory.  This is because these transformations are a
subgroup of ${\rm Sp}(4 + 2 N_f, {\bf R})$ which leaves invariant the
K\"ahler potential with the spurion superfields.  The microscopic
duality (\ref{micros}) induces a transformation on the dilaton spurion
superfield that generally does not leave invariant the K\"ahler
potential.  We can see from (\ref{micros}) and ({\ref{duals}) that
\be
{\rm Im} ({\cal T}_{0,D}^{\Gamma} {\overline {\cal T}_0^{\Gamma}})  
\not=
{\rm Im} ({\cal T}_{0,D} {\overline {\cal T}_0} ).
\label{noinvt}
\ee
On the other hand, the K\"ahler potential for the mass spurions  
satisfies
\be
{\rm Im} ({\cal M}_D^{f,\Gamma} {\overline {\cal M}_f^{\Gamma}}) = 
{\rm Im} ({\cal M}_D^f {\overline {\cal M}_f} ),
\ee
{\it i.e.} in the softly broken theory, 
the microscopic duality symmetry is lost for the dilaton spurion
but it is preserved for the mass spurion.  More precisely, the  
``cosmological" term
associated to the mass spurions
\be
V_{\rm eff}^{(VM)} = \left( {b_{fa}b_{ag} \over b_{aa}} - b_{fg}  
\right)
\left( {1 \over 2}D_f D_g + {\overline F}_f F_g \right)
\ee
is invariant if at the same time we also transform
\be
(F_f, D_f) \rightarrow t_{fg} (F_g, D_g)
\ee
But if $\g \not=0$ in (\ref{micros}), one can check
that the cosmological term is not invariant if $F_0$ or $D_0$ are not  
zero.

\subsection{Expressions for the dilaton spurion couplings.} 
\indent

The next step to study the behavior of the softly broken models is to
obtain explicit expressions for the dilaton and mass couplings in
terms of quantities associated to the Seiberg-Witten curve of the
corresponding theory. In this subsection we consider the dilaton
spurion couplings.

In the case of the self-dual theories, we cannot use the Euler
relation for the prepotential in order to extract the dependence on
$\tau_0$, as this is now a truly dimensionless variable. Therefore, in
order to compute the dual spurion $\tau_{0, D}$ (which involves
essentially the RG equation for this theory), we have to use the
Riemann bilinear relations, as it was done in \cite{amz} to compute
the dual masses.

First of all we have to
compute the derivatives of the Seiberg-Witten differential with
respect to the coupling $\tau_0$ and the masses. After a somewhat  
long
computation, we find 
\be \Bigl( {\partial \lambda_{SW} \over \partial
\tau_0} \Bigr)_{\tilde u} = -{{\sqrt 2} \over 8 \pi} {dx \over y}
\biggl( {(\a - \b)' \over (\a-\b)} \tilde u + \pi i (x + { R\over 3}
\a \b) \biggr),
\label{ltau}
\ee
where the prime denotes the derivative w.r.t. $\tau_0$. It is easy to  
check that this is an abelian 
differential of the second kind, with a double pole at $x =\infty$. 
To obtain the expression (\ref{ltau}), one must use the identities
\bea
\a' = {\pi i \over 3} \alpha (E_2 (\tau_0) + 2 \b - \a), \nonumber\\
\b' = {\pi i \over 3} \alpha (E_2 (\tau_0) + 2 \a - \b).  
\label{abders}
\eea
With this result we can already compute the dual spurions, using the  
strategy introduced in 
\cite{amz}. Taking $a$, $\tau_0$ and $m_f$ as the independent  
variables of the prepotential, we have the relation  
\be
\Big({\pa\tau_{0,D} \over \pa a}\Big)_{\tau_0} = \Big({\pa a_D \over  
\pa \tau_0}\Big)_a
 = \oint_{\a_1} \Big({\pa \lambda_{SW} \over \pa \tau_0 }\Big)_a.
\ee
We then consider the vanishing $(2, 0)$-form
$$\Big({\pa \lambda_{SW} \over \pa a }\Big)_{\tau_0} \wedge
\Big({\pa \lambda_{SW} \over \pa \tau_0}\Big)_a.$$
Notice that the first form is holomorphic and can be written as
\be 
\psi= \Big({\pa\lambda_{SW} \over \pa a}\Big)_{\tau_0} = {{\sqrt 2}  
\over 8 \pi} {dx \over y} \Big({\pa {\tilde u} \over \pa  
a}\Big)_{\tau_0},
\label{one}
\ee
and the second form is 
\be 
\eta=\Big({\pa \lambda_{SW} \over \pa \tau_0}\Big)_a =  \Big( {\pa  
\lambda_{SW}  
\over \pa {\tilde u} } \Big)_{\tau_0}  \Big( {\pa {\tilde u} \over  
\pa \tau_0} \Big)_{\tau_0} + \Big( {\pa \lambda_{SW} \over \pa  
\tau_0}  
\Big)_{\tilde u} . 
\label{two}
\ee
The first piece of  (\ref{two}) is holomorphic, and the second part  
is given in (\ref{ltau}). The Riemann bilinear relation gives 
\be
\label{birel}
\oint_{\a_1} \Big({\pa \lambda_{SW} \over \pa \tau_0}\Big)_a =
2 \pi i {\rm Res}_{x=\infty} (\pi \cdot \eta), 
\label{riemann}
\ee
where $\pi (s) = \int_{s_0}^s \psi$. We have taken into account that
\bea
\oint_{\a_2} \Big({\pa \lambda_{SW} \over \pa \tau_0 }\Big)_a &=&  
\Big({\pa \over
 \pa \tau_0}\Big)_a
\oint_{\a_2} \lambda_{SW} = 0,
\\
\oint_{\a_2} \Big( {\pa \lambda_{SW} \over \pa a }\Big)_{\tau_0} &=&
 \Big({\pa \over \pa a }\Big)_{\tau_0}
\oint_{\a_2} \lambda_{SW} = 1.
\eea
The residue is easily computed using for instance the uniformization  
of the curve. The pole 
is at $z=0$ in the ${\bf  C} /\Lambda$ plane, and we finally obtain
\be
\tau^{0a}= {1 \over 4}   \Big( {\pa {\tilde u} \over \pa a}  
\Big)_{\tau_0}. 
\label{monpar}
\ee

Integrating w.r.t. to $a$, we can read the RG equation for the  
$N_f=4$  
theory, which gives the dual dilaton spurion:
\be 
\tau_{0,D}= {\partial {\cal F} \over \partial \tau_{0}}= {1 \over 4}  
{\tilde u}.
\label{rgeq}
\ee
As ${\tilde u}$ is a modular form of weight  $(2,0)$, this is in  
perfect agreement  with the second 
equation in (\ref{duals}). Another check of this expression can be  
obtained if one considers 
the massless case, where $m_f=0$ for all $f=1, \cdots, 4$.  
Then we have for the 
prepotential and the periods the simple expressions  
(\ref{semiperiods}), (\ref{semiprep}), and we can see that the above  
relation 
is trivially satisfied.  The RG equation for the mass-deformed ${\cal  
N}=4$  
theories with gauge 
group $SU(N)$ has been obtained using the Calogero-Moser system in  
\cite{calo}. Related 
considerations have been made in \cite{inst}.  

Now we want to compute the second derivative of the prepotential  
w.r.t. the microscopic 
coupling. From (\ref{rgeq}) we find 
\be
\tau_{00} = {\partial^2 {\cal F} \over \partial \tau_{0}^2}= {1 \over  
4} \Big( {\partial \tilde u \over \partial \tau_0}  \Big)_{a}=
-{1 \over 4} { (\partial a / \partial  \tau_0)_{\tilde u} \over  
(\partial a / \partial  
{\tilde u})_{\tau_0}}. 
\label{tausq}
\ee

To obtain an explicit expression for $(\partial a / \partial   
\tau_0)_{\tilde u}$, we use again  uniformization of the  
Seiberg-Witten curve. In this way we obtain
\bea
 \Bigl( {\partial \lambda_{SW} \over \partial \tau_0} \Bigr)_{\tilde  
u} &=& -{{\sqrt 2} \over 8 \pi} 
dz  \biggl( {(\a - \b)' \over (\a-\b)}  + {\pi i \over 3} (\alpha  
+\beta) \biggr) {\tilde u}   - { i {\sqrt 2} \over 2} \wp (z)  
dz\nonumber\\
& +&  { {\sqrt 2}  i\over  8  } dz  {R \over 18} \bigl[ (\alpha +  
\beta)^2 -3 (\alpha - \beta)^2 -6 \alpha \beta \bigr]\nonumber\\
&=&- i {{\sqrt 2} \over 2} dz \biggl[ {1 \over 12} E_2 (\tau_0)  
{\tilde u} + \wp(z) + {R \over 36} E_4 (\tau_0) \biggr],
\label{explicito}
\eea
where we have used (\ref{efour}) and (\ref{abders}). We finally find
\be
\tau^{00}= { i \pi \over 12} \biggl[ E_2(\tau_0){\tilde u} + {R \over  
3} E_4 (\tau_0) \biggr] 
-{ 2\pi i  \over \omega} \zeta ( {\omega \over 2}), 
\label{taunutnut}
\ee
where $\omega$ is the period corresponding to the $a$ variable. We  
can write this expression in another way using that 
\be
{8 \pi i \over \omega}  \zeta ( {\omega \over 2})= { \pi i \over  
24} E_2 (\tau^{aa}) \biggl( { 
d{\tilde u} \over d a }\biggr) ^2,
\label{etwoid}
\ee
to obtain
\be
 \tau^{00}= { i \pi \over 4} \biggl[ E_2(\tau_0){\tilde u \over 3}-  
{1 \over 24} E_2 (\tau^{aa})\Big( { 
d{\tilde u} \over d a} \Big) ^2  \biggr] + {i \pi R \over 36} E_4  
(\tau_0). 
\label{taunutbis}
\ee

Using this expression and the properties of the Eisenstein series  
under modular 
transformations,
\bea
E_2 (\tau_0) & \rightarrow & (\g \tau_0 + \d)^2 \biggl( E_2(\tau_0) +  
{ 12 \g \over 2 \pi i ( \g \tau_0 
+ \d)} \biggr), \nonumber\\
E_4 (\tau_0) & \rightarrow &  (\g \tau_0 + \d)^4 E_4 (\tau_0), 
\label{eismod}
\eea
we can explicitly check the microscopic duality transformation in  
(\ref{monotrans}) for $\tau^{00}$ \footnote{The structure of this  
coupling has been obtained in a different context 
in \cite{mw}.}.

\subsection{Expressions for the mass spurion couplings.} 
\indent

The dual mass in the $SU(2)$ theories is given by the general  
expression \cite{amz}
\be
m_D^f= \sum_{n=1}^{N_p}S_n^f \int_{x_n^-}^{x_n^+} \lambda_{SW},
\label{dualmass}
\ee
where $x_n^{\pm}$ denote the two roots of $y$ corresponding to the  
position of the poles. 
An explicit expression can be obtained using uniformization  
\cite{amz}. In the $N_f=4$ case 
we have:
\bea
m_D^f &=& - { {\sqrt 2} \over 8 \pi}  \biggl(  2({\tilde u} -e_1 R)  
z_f+ i \sum_{g \not= f} 
m_g [ 2z_f \zeta (z_g) - {\rm log} { \sigma ( z_f-z_g) \over \sigma  
(z_f + z_g) } ] \nonumber\\
&+& im_f [ 2z_f \zeta(z_f) + {\rm log} \sigma(2z_f)]\biggr). 
\label{dualmassfour}
\eea

To obtain the mass spurion couplings, we must compute the derivative  
of the Seiberg-Witten differential w.r.t. to the 
masses. One finds,
\be
\Bigl( {\partial \lambda_{SW} \over \partial m_f} \Bigr)_{\tilde u} =  
-{{\sqrt 2} \over 8 \pi} {dx \over 
y} \biggl( m_fe_1 + {q_f  \over x + \alpha \beta m_f^2} \biggr),
\label{lmass}
\ee
and after integrating on the one-cycles one obtains the derivatives  
of the periods w.r.t. 
the masses (keeping ${\tilde u}$ fixed). 
The mass couplings are then given by:
\bea 
\tau^{af} &=& \Bigl( {\partial a_D \over \partial m_f} \Bigr)_{\tilde  
u} - 
\tau^{aa} \Bigl( {\partial a_D \over \partial m_f} \Bigr)_{\tilde u} 
= -{1 \over \omega} z_f, 
\\
\tau^{0f}&=& -{1 \over 4} \Bigl( {\partial {\tilde u}  \over \partial  
a} \Bigr)_{m_f}\Bigl( {\partial a \over \partial m_f} \Bigr)_{\tilde  
u} \nonumber\\
&=& { { \sqrt 2} \over 4} m_f e_1 - { i {\sqrt 2}  \over 2} [  
\zeta(z_f) - { 2 z_f \over \omega} 
\zeta ({\omega \over 2})] ,\\
\tau^{fg} &=& - {{\sqrt 2} \over 2} \int_{x_f^-} ^{x_f^+} \Bigl[  
\Bigl( {\partial \lambda_{SW}  \over \partial {\tilde u} } \Bigr)_{  
m_f } \Bigl( {\partial {\tilde u}  \over \partial m_g} \Bigr)_a +  
\Bigl( {\partial \lambda_{SW} \over \partial m_g} \Bigr)_{\tilde u}  
\bigr] \nonumber\\
&=& { i \over \pi} (1-\delta_{fg}) \biggl( {1 \over 4} {\rm log}  {  
\sigma ( z_f-z_g) \over \sigma (z_f + z_g) } - 
{z_f z_g\over \omega }  \zeta({ \omega \over 2}) \biggr)  
\nonumber \\
&-&  { i \over \pi}\delta_{fg}  \biggl( {1 \over 4} {\rm log}    
\sigma ( 2z_f)  
+{z_f ^2\over \omega }  \zeta({ \omega \over 2}) \biggr).
\label{masscouplings}
\eea


\bigskip
 
\section{Extension to Higher Rank Gauge Groups}
\setcounter{equation}{0}
\indent

${\cal N}=2$ supersymmetric QCD theories with a vanishing one-loop
beta function are supposed to enjoy exact conformal invariance when
the hypermultiplets are massless, and to have a duality invariance in
the massive case. Explicit curves showing this invariance have been
constructed in \cite{donwit} for the mass-deformed $SU(N)$ ${\cal
N}=4$ theories, and in \cite{aps, as, ho, han, mn} for the classical
gauge groups.

Most of the considerations explained above extend 
in a straightforward way to the case of gauge groups of rank $r$. An  
important difference is 
that the microscopic duality group will be, in general, a semidirect  
product of a subgroup $M$ of $SL(2, {\bf Z})$ with a group which acts  
on the masses. A microscopic duality transformation  corresponding to  
the matrix 
\be
\Gamma =  \left(\begin{array}{cc}  \alpha &  \beta  \\  \gamma &  
\delta \end{array}\right) \in 
M 
\label{mmatrix}
\ee
together with an action on the masses $m_f \rightarrow t_{fg}m_g$,  
will induce in general an inhomogeneous $Sp(2r, {\bf Z})$  
transformation acting on $(a_D^I, a_I)$: 
\be 
\left(\begin{array}{c} a_D^I \\ a_I \end{array}\right) \rightarrow   
\left(\begin{array}{c}  A^I_{~J}a_D^J + B^{IJ}a_J + p^{If} m_f/{\sqrt  
2}  \\  C_{IJ} a_D^J + D_I^{~J} a_J  +q_I^fm_f/{\sqrt 2}  
\end{array}\right) .
\label{simplec}
\ee
The duality transformations for the couplings generalize in a  
straightforward way. The effective 
$SL(2, {\bf Z})$ matrix is now a symplectic one, while the modular  
factors involving 
$\tau_0$ remain the same. In particular, the dual spurion is still a  
modular form of weight 
two with respect to the modular subgroup $M \subset SL(2, {\bf Z})$. 

In fact, we can verify this statement and compute the explicit form  
of the dual spurion 
for all the classical gauge groups, using the results of  
\cite{aps,as}. This is a nontrivial 
check of the modular transformations we have derived using only the  
properties of the 
prepotential. It also gives the RG equation for the self-dual ${\cal  
N}=2$  
theories with 
classical groups, extending the results of \cite{matone}. 

\bigskip

a) $SU(r+1)$ with $N_f=2r+2$ hypermultiplets in the fundamental  
representation. 

The curve is given by
\be
y^2=P^2(x) + 4h(h+1) \prod_{j=1}^{2r+2} (x-m_j-2 \mu h),
\label{suncurve}
\ee
where $m_j$, $j=1, \cdots, 2r+2$ are the masses of the   
hypermultiplets, and 
\bea
P(x)= \prod_{a=1}^r (x- \phi_a)= x^r-\sum_{k=2}^{r+1}u_k  
x^{r+1-k},\nonumber\\
\mu={1 \over N_f} \sum_{i=1}^{N_f} m_f, \,\,\,\,\,\ h(\tau_0)= {  
\beta- \alpha \over \alpha-2 \beta}.
\label{suncuan}
\eea
The Seiberg-Witten differential can be written as 
\be
\lambda_{SW}={x-2 \mu h \over 2 \pi i} d {\rm log} \biggl[ {P(x)-y  
\over P(x)+y} \biggr].
\label{sundiff}
\ee
For $r>1$, this curve is invariant under the duality group  
$\Gamma^{0} (2)$, generated by 
$T^2$ and $S$, $m_j \rightarrow m_j-2 \mu$. Notice that the Casimirs  
are modular forms of 
weight zero. One can easily obtain from (\ref{sundiff}) the  
expression
\be
\Bigl( {\partial \lambda_{SW} \over \partial \tau_0} \Bigr)_{u_k}= 
{1 \over 2 \pi i} { dx \over y} \biggl[ 4 \mu h' {dP \over dx} -  
{(h(h+1))' \over h(h+1)} P(x) \biggr],
\label{suntau}
\ee
where the $'$ indicates derivative w.r.t. $\tau_0$. This is a  
differential of the second kind with a 
double pole at $x = \infty$ (one can check that the residue  
vanishes). The basis of holomorphic 
differentials on the hyperelliptic curve is the usual one, $\omega_k  
= x^{r+1-k}(dx/y)$, and 
$(\partial \lambda_{SW}/ \partial u_k) = -(1/\pi i ) \omega_k$. Using  
a straightforward generalization of the argument in section 3.2.1, we  
find
\be
\tau_{0,D} ={ \partial {\cal F} \over \partial \tau_0} = (\alpha-2  
\beta) u_2.
\label{suntaud}
\ee
Notice that $\alpha - 2 \beta$ is a modular form for $\Gamma^0(2)$ of  
weight two. 

\bigskip
b) $Sp(2r)$ with $N_f=2r+2$ hypermutliplets in the fundamental  
representation. 

The curve is 
\be
y^2=x P^2(x) + 2gQ P(x) +g^2 R,
\label{spcurve}
\ee
where  
\bea
P(x)= \prod_{a=1}^r (x- \phi_a^2)= x^r-\sum_{k=1}^{r}u_k  
x^{r-k},\nonumber\\
Q=\prod_{j=1}^{2r+2}m_j, \,\,\,\,\,\ Q^2-xR= \prod_{j=1}^{2r+2}  
(x-m_j^2), \,\,\,\,\ g(\tau_0)= { \alpha- \beta \over \alpha+ \beta}.
\label{spcuan}
\eea
The Seiberg-Witten differential is given by 
\be
\lambda_{SW}={{\sqrt x} \over 2 \pi i} d {\rm log} \biggl[ {xP(x)  
+gQ-{\sqrt x}y \over xP(x)+gQ+{\sqrt x}y} \biggr].
\label{spdiff}
\ee
For $r>1$, the duality group is now $\Gamma_{0} (2)$, with generators 
$T$, $\prod_j m_j\rightarrow -\prod_j m_j$, and $ST^2S$. We find 
\be 
\Bigl( {\partial \lambda_{SW} \over \partial \tau_0} \Bigr)_{u_k}= 
-{1 \over 2 \pi i} { dx \over y} {g' \over g}P(x),
\label{sptau}
\ee
and the dual spurion turns out to be
\be
\tau_{0,D} ={ \partial {\cal F} \over \partial \tau_0} =-  
(\alpha+\beta) u_1,
\label{sptaud}
\ee
which is again a modular form of weight two with respect to  
$\Gamma_0(2)$. 

\bigskip
c) $SO(2r+1)$ with $N_f=2r-1$ vector hypermultiplets and $SO(2r)$  
with $N_f=2r-2$ 
vector hypermultiplets. 

The curve can be written in both cases in the form
\be
y^2=xP^2(x) + 4f A(x), 
\label{socurve}
\ee
where $P(x)$ is given by the expression in (\ref{spcuan}), $f=h(h+1)$  
with $h$ given in (\ref{suncuan}), and 
\bea
A(x)&=&x^2\prod_{j=1}^{2r-1} (x-m_j^2), \,\,\,\,\,\ {\rm for} \,\  
SO(2r+1), \nonumber\\
&=& x^3\prod_{j=1}^{2r-2} (x-m_j^2), \,\,\,\,\,\ {\rm for} \,\  
SO(2r).
\label{socuan}
\eea
The duality group is again $\Gamma^0(2)$, with no action on the  
masses. 
The Seiberg-Witten differential is given in both cases by
\be
\lambda_{SW}={{\sqrt x} \over 2 \pi i} d {\rm log} \biggl[  
{xP(x)-{\sqrt x}y \over xP(x)+{\sqrt x}y} \biggr].
\label{sodiff}
\ee
We easily find 
\be 
\Bigl( {\partial \lambda_{SW} \over \partial \tau_0} \Bigr)_{u_k}= 
-{1 \over 4 \pi i} { dx \over y} {f' \over f}P(x),
\label{sotau}
\ee
and the dual spurion is
\be
\tau_{0,D} ={ \partial {\cal F} \over \partial \tau_0} =  
(\alpha-2\beta) u_1.
\label{sotaud}
\ee
In all these cases, the spurion coupling $\tau_{00}$ has the  
structure 
\be
\tau_{00}={ \pi i \over 3} E_2 (\tau_0) \tau_{0,D} -  
f_1(\tau_0)\sum_{I=1}^r \Bigl( {\partial 
u \over \partial a^I} \Bigr)_{\tau_0}  \Bigl( {\partial 
a^I\over \partial \tau_0} \Bigr)_{u_k}
+ {\pi i \over 3} f_2(\tau_0) u ,
\label{hrdualc}
\ee
where $u$ denotes the relevant Casimir ($u_2$ for $SU(N_c)$, $u_1$  
for the other 
classical groups), and $f_1(\tau_0)$, $f_2 (\tau_0)$ are modular  
forms of weight 
two with respect to the relevant duality group (which can be  
explicitly computed from 
the above expressions). In principle, one can compute the  
derivatives of the 
Seiberg-Witten periods $a^I$ w.r.t. the coupling $\tau_0$ by  
integrating the expressions 
(\ref{suntau}), (\ref{sptau}) and (\ref{sotau}) along the  
corresponding homology cycles. 
Explicit expressions in terms of derivatives of theta-functions can  
be possibly obtained using 
the techniques of \cite{calo,gorsky}.

 
\bigskip

\section{The Mass-Deformed ${\cal N}=4$ Theory}  
\setcounter{equation}{0}

\subsection{Structure of the Seiberg-Witten solution.}
\indent

As we have already mentioned, the curve for the mass-deformed ${\cal
N}=4$ theory is obtained from (\ref{curva}) by simply choosing $m_1 =
m_2 = m/2$, $m_3=m_4=0$.  Also, the normalization of the
Seiberg-Witten differential changes and we have instead a factor of
$1/4 \pi$ in the r.h.s. of (\ref{swabel}).  As there is only a mass
parameter, the theory is much simpler to study, although we loose the
triality part of the duality group which acts on the masses.  The most
important advantage is that one can easily find an explicit expression
for the effective roots $\hat e_i$ of the curve in the Weierstrass
form (\ref{weier}):
\be
\hat e_i = {1 \over 4} e_i \tilde u- {1 \over 16} m^2 \biggl( { 2  
E_4 (\tau_0)  \over 3} - e_i^2 \biggr), 
\,\,\,\,\, i = 1, \cdots, 3, 
\label{raizes}
\ee

A much more compact expression 
can be written for the position of the pole:
\be
\wp (z_0) = \hat e_1 - {\alpha \beta \over 16} m^2.
\label{defpolo}
\ee

The expression for the periods is then,
\be
a_i = {{\sqrt 2} \over 2 \pi} \biggl( (\tilde u- e_1 {m^2 \over 4}) 
\omega_i + i  
m [ \omega _i \zeta (z_0) - 2 z_0 \zeta ({\omega_i \over 2})]  
\biggr),  
\,\,\,\,\,\,\ i =1,2,
\label{nfourperiodos} 
\ee

In this parametrization, the effective low energy theory with modulus
$\tau^{aa}(\tilde u, \tau_0, m)= {\pa a_2}/{\pa a_1}$ approaches the
microscopic theory with modulus $\tau_0$ in the massless limit (or
weak coupling region) $m^2/\tilde u \rightarrow 0$.  In this way, we
identify $\sqrt{2}|a_1|$ with the mass of the electric $W$-bosons,
$\sqrt{2}|a_1 \pm m/\sqrt{2}|$ with the mass of the components
$\sigma^1$ and $\sigma^2$ of the adjoint quark, and $|m|$ with the
mass of the component $\sigma^3$.  This theory has an stable BPS
spectrum $(n_m, n_e)$ for relatively prime magnetic and electric
quantum numbers ($n_m$ and $n_e$, respectively) \cite{ferrari}. Their
masses are given by $\sqrt{2}|n_m a_2 + n_e a_1 + S m/\sqrt{2}|$, for
baryon number $S$.  There are three points at the $\tilde u$-plane
where the curve describing the low energy effective theory
degenerates.  They are given by ${\tilde u}_i = e_i m^2/4$, $i=1, 2,
3$.  At these singularities, the effective coupling constant has a
logarithmic singularity, $\tau_{aa}^{(i)} \sim {\rm ln}(\tilde u -
\tilde u_i)$, due to a massless charged particle in the infrared
limit.  If we use the macroscopic frame $a = a_1$ and $a_D = a_2$,
given by (\ref{nfourperiodos}), the numerical analysis gives that
$a_1({\t u}_1) = a_2({\t u}_2) = a_2({\t u}_3) + a_1({\t u}_3) = 0$.
It means that in this frame, the massless particles are: an adjoint
quark $(0, 1)$ at ${\t u}_1$, a monopole $(1, 0)$ at ${\t u}_2$ and a
dyon $(1, 1)$ at ${\t u}_3$.  Notice that the monodromy base point has
positive imaginary part (see fig. 1).

\subsection{Microscopic duality.}
\indent

Using standard properties of elliptic functions, one can check that
the transformation properties of the Seiberg-Witten periods and the
couplings under a microscopic duality transformation are indeed given
by the expressions (\ref{macros}) and (\ref{monotrans}). From the
analytic expressions of the effective roots $e_i(\tilde u, \tau_0, m)$
in (\ref{raizes}), we observe that a microscopic duality
transformation $\Gamma$ induces the same permutation on the effective
roots (but with weight four) as on the microscopic ones.  This means
that in the macroscopic duality frame given by the formulae
(\ref{raizes}-\ref{nfourperiodos}), $\Gamma$ gives the same effective
duality transformation on the low energy theory, up to a monodromy
transformation \cite{ferrari}.

We consider now in some detail the
action of the generators $S$ and $T$.
Under the interchange $\tau_0 \rightarrow -1/\tau_0$, 
the microscopic roots $e_1$ and $e_2$ are permuted (with an  
additional $\tau_0^2$ factor).
This has the following effect on the Seiberg-Witten singularities:
\be
\tilde u_1\left(-{1 \over \tau_0}, m \right) = \tau_0^2 \tilde  
u_2(\tau_0, m), \,\,\,\,\ 
\tilde u_3\left(-{1 \over \tau_0}, m \right) = \tau_0^2 \tilde  
u_3(\tau_0, m).
\ee
From the expressions (\ref{raizes}), we have that 
\bea 
e_1 \left(\tilde u, -{1 \over \tau_0}, m \right) = 
\tau_0^4 e_2 \left({\tilde u \over \tau_0^2}, \tau_0, m \right),  
\nonumber
\\
e_3 \left(\tilde u, -{1 \over \tau_0}, m \right) = 
\tau_0^4 e_3\left({\tilde u \over \tau_0^2}, \tau_0, m \right),
\label{Seff}
\eea
and this gives the following transformation of the effective periods  
$(\omega_1, \omega_2)$ 
in (\ref{periodos}), 
\bea
\omega_1\left(\tilde u, -{1 \over \tau_0}, m\right) &=& 
- \tau_0^{-2} \omega_2\left({\tilde u \over \tau_0^2}, \tau_0,  
m\right),
\nonumber
\\
\omega_2\left(\tilde u, -{1 \over \tau_0}, m\right) &=& 
\tau_0^{-2} \omega_1\left({\tilde u \over \tau_0^2}, \tau_0, m\right)  
\,.
\eea
Less obvious is the  
$S$-duality transformation of $z_0$, since it picks a shift.
From (\ref{defpolo}), we have that
\be
{1 \over \tau_0^4} \wp 
\left( z_0\left(\tilde u, -{1 \over \tau_0}\right) \right) =
{\hat e}_2\left({\tilde u \over \tau_0^2}, \tau_0\right) 
- {1 \over 16} \alpha \left(\alpha- \beta \right) m^2.
\label{spolo}
\ee
We can now use the properties of the Weierstrass function \cite{akhi} 
\bea
\wp(\lambda z;\{\lambda \omega_1, \lambda \omega_2\}) &=& 
\lambda^{-2} \wp(z; \{\omega_1, \omega_2 \}),
\nonumber \\ 
 \wp (z+ {\omega_i \over 2})&=& \hat e_i + { (\hat e_i-\hat e_j)(\hat  
e_i -\hat e_k) \over \wp(z)-\hat e_i} 
\label{wptrans}
\eea
to see that 
\bea
\tau_0^2 z_0\left(\tilde u, -{1 \over \tau_0}\right) &=& 
z_0\left({\tilde u \over \tau_0^2}, \tau_0\right) 
+ {1 \over 2} \omega_3\left({\tilde u \over \tau_0^2}, \tau_0\right) 
\nonumber\\
&+& n_1\omega_1\left({\tilde u \over \tau_0^2}, \tau_0\right) 
+ n_2 \omega_2\left({\tilde u \over \tau_0^2}, \tau_0\right),
\label{spoloex}
\eea
where $n_i$ are arbitrary integers which arise because of the  
periodicity of the 
Weierstrass functions. It was shown in \cite{amz,ferrari,bf} that  
these ambiguity 
corresponds precisely to the inhomogeneous piece of 
the macroscopic duality transformations,
which physically means a redefinition of the baryon charge. 
Taking all of this  
into account, and using Legendre's relation, we easily find the  
$S$-duality relation:
\bea
a_1 \left(\tilde u, -{1 \over \tau_0}, m\right) &=& 
- a_2\left({\tilde u \over \tau_0^2}, \tau_0, m\right)
+ \left(1 + 2 n_1\right) {m \over \sqrt{2}},
\nonumber
\\
a_2\left(\tilde u, -{1 \over \tau_0}, m\right) &=& 
a_1\left({\tilde u \over \tau_0^2}, \tau_0, m\right) 
+ \left(1 + 2 n_2\right) {m \over \sqrt{2}},
\label{Saeff}
\eea
\figalign{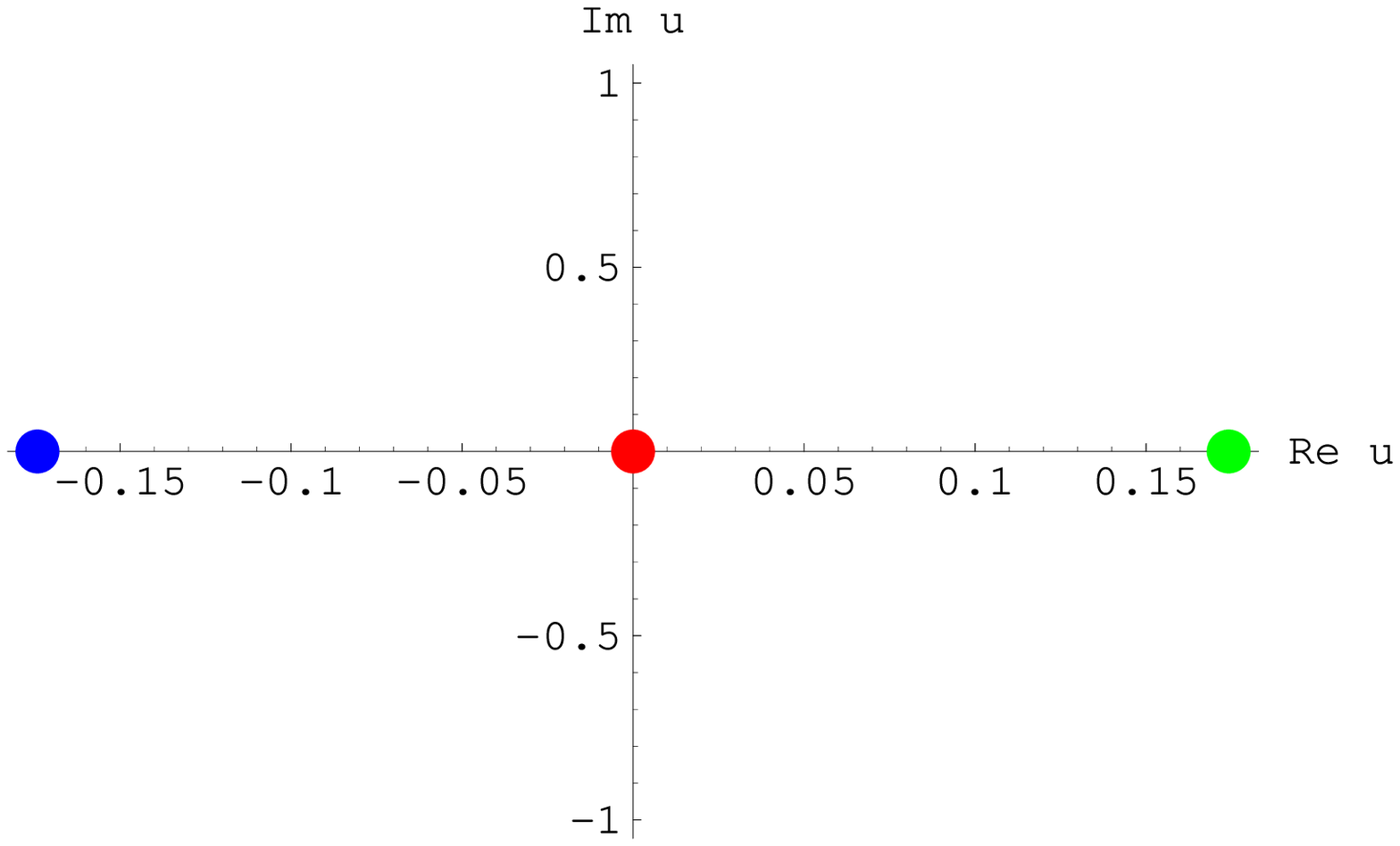}{\pu}{The Seiberg-Witten singularities at the  
$\tilde  
u$-plane 
for microscopic coupling $\tau_0 = i$. From left to right, we have: 
$\tilde u_2$ (monopole), $\tilde u_3$ (dyon) and $\tilde u_1$  
(adjoint quark).}
{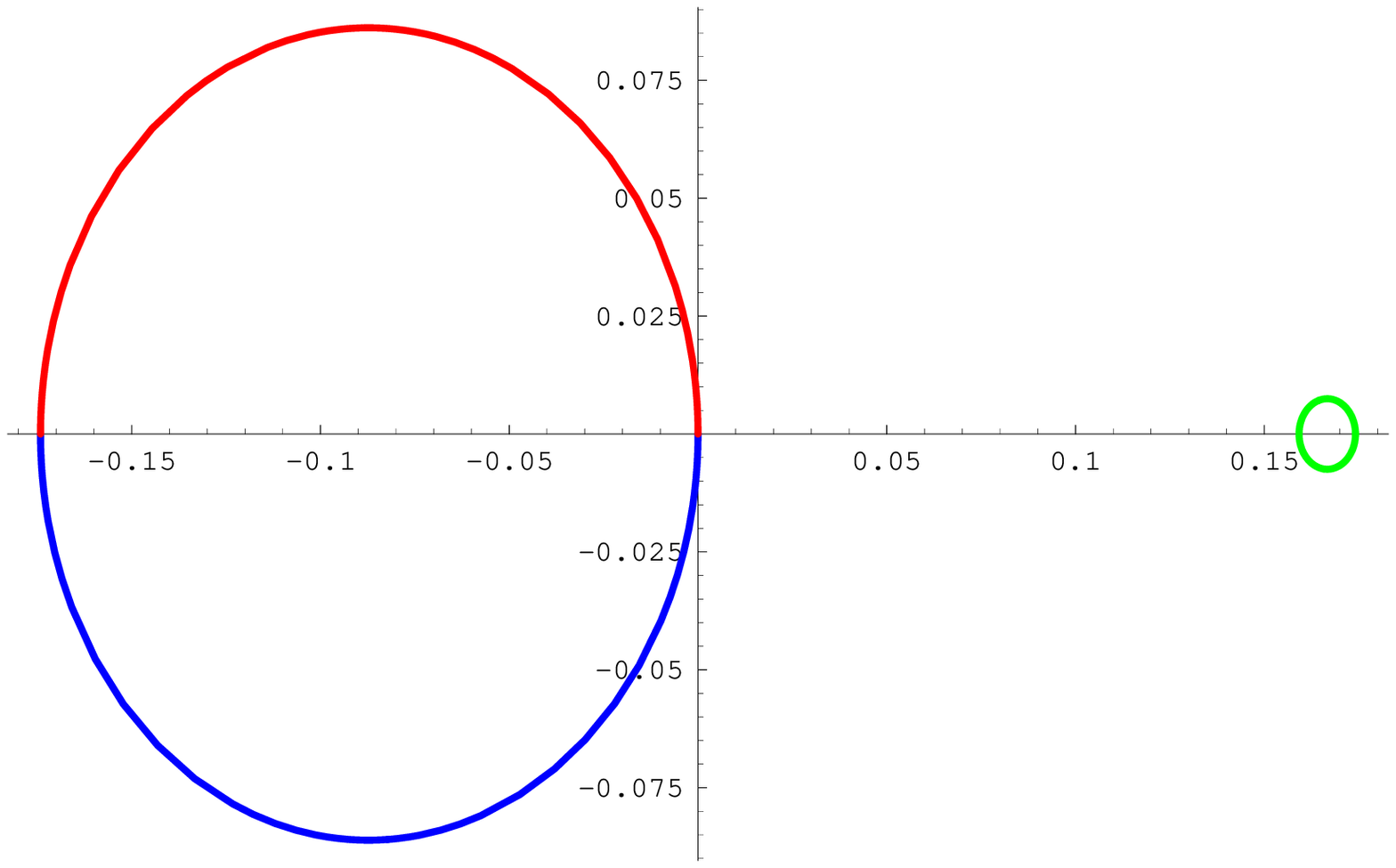}{\puti}{The counterclokwise rotation of the  
Seiberg-Witten  
singularities 
under the shift $\tau_0 = i$ to $\tau_0 = i + 1$.} 

Notice that in the frame $a=a_1$ and $a_D=a_2$, the macroscopic
$S$-duality transformation coincides with the microscopic one (up to
inhomogeneous terms and monodromy transformations, which have been
found in \cite{ferrari}).  This is because in the massless limit, the
Seiberg-Witten periods become (\ref{semiperiods}). For another
macroscopic frame $(a, a_D)$, related to the variables (\ref{swp}) by
a general $SL(2,{\bf Z})$ transformation, the macroscopic duality
transformation will be different.

We emphasize that $S$-duality is a symmetry of the theory: the theory
at the vacuum labeled by $\tilde u$ and coupling $-1/\tau_0$ is {\it
equivalent} to the theory at the vacuum labeled by $\tilde u/\tau_0^2$
and coupling $\tau_0$, but described in a macroscopic $S$-dual frame.
For instance, in \fig\pu\ we show the position of the Seiberg-Witten
singularities for $\tau_0=i$.  An $S$-duality transformation will map
the theory at $\tilde u$ to the one at $-\tilde u$, with the
singularities $\tilde u_1$ and $\tilde u_2$ rotated. We see that this
corresponds to a $\pi$ rotation of the $\tilde u$-plane, since in this
case $\tau_0=i$ is a fixed point under $S$-duality. For other values
of the microscopic coupling, the action will be more complicated.

Under the interchange $\tau_0 \rightarrow \tau_0 + 1$, the microscopic
roots $e_2$ and $e_3$ are permuted (without any modular factor). This
gives
\be
\tilde u_1(\tau_0 +1, m) = \tilde u_1 (\tau_0, m)\ ,
\ \  
\tilde u_2(\tau_0 \pm 1, m) = \tilde u_3 (\tau_0, m)
\ee
The effective roots are permuted in the same way.
The pole $z_0$ is invariant under $T$-duality, 
and one can easily derive the relation:
\bea
a_1(\tilde u, \tau_0 + 1, m) &=& a_1(\tilde u, \tau_0, m),
\nonumber
\\
a_2(\tilde u, \tau_0 + 1, m) &=& a_2(\tilde u, \tau_0, m) +  
a_1(\tilde u, \tau_0, m).
\label{Taeff}
\eea
In \fig\puti\
we show the movement of the singularities in the $\tilde u$-plane  
when $\tau_0=i$, under the shift 
$\tau_0 \rightarrow \tau_0 +1$. We observe that $\tilde u_2$ and  
$\tilde u_3$
rotate counterclockwise. 
One can check that in the $T$-dual situation, the massless particle
at $u_3$ becomes a dyon $(1, -1)$, 
As $\theta_{\rm eff}(\tilde u, \tau_0 + 1) =  
\theta_{\rm eff}(\tilde u, \tau_0) +1$, by the Witten effect
the physical electric charges in the $T$-dual frame are the same.


\bigskip

\section{Soft Breaking of the Mass-Deformed 
${\cal N}=4$ Super Yang-Mills Down to ${\cal N}=0$}
\setcounter{equation}{0}
\indent

In this section we study in detail the vacuum structure of the softly
broken, mass-deformed ${\cal N}=4$ super Yang-Mills with gauge group
$SU(2)$.  We will only consider the soft breaking with a non zero $F$
term for the dilaton spurion (its absolute value, denoted by $F_0$,
will be the supersymmety breaking scale).  Since we will work with an
adjoint mass $m$ different from zero, in the numerical analysis we
define our units in such a way that $m=1$.

The couplings for the mass-deformed ${\cal N}=4$ theory  
can be obtained from the 
ones for the $N_f=4$ theory with masses $(m/2, m/2, 0,0)$, as the  
Seiberg-Witten curve is 
the same, but one has to take into account the different 
normalizations. We have for instance,
\bea
\tau_{0,D}&=& {\tilde u},\nonumber\\
\tau^{00}&=&  i \pi  \biggl[ E_2(\tau_0){\tilde u \over 3}- {1 \over  
6} E_2 (\tau^{aa})\Big( { 
d{\tilde u} \over d a} \Big) ^2  \biggr] + {i \pi m^2 \over 36} E_4  
(\tau_0).
\label{changes}
\eea

\subsection{General properties.}
\indent

The ${\cal N}=2$ soft breaking method preserves the ${\cal N}=2$
supersymmetric expression of the Lagrangian, with the only difference
that some microscopic couplings are promoted to ${\cal N}=2$ spurion
superfields. Seiberg and Witten \cite{swone, swtwo} obtained the  
exact
holomorphic part of the ${\cal N}=2$ effective Lagrangian, where the
massive particles have been integrated out.  The whole effective
Lagrangian has an infinite series of terms, which are naturally
expanded in inverse powers of the Wilsonian ultra-violet (UV) cut-off
$\mu$ This scale is set by the mass of the lightest particle that has
been integrated out. Notice that this mass is given by the
Seiberg-Witten solution, since the lightest particle is a BPS
state. For instance, in the pure ${\cal N}=2$ theory \cite{swone},
when the vacuum is very close to the monopole singularity
$u_{\Lambda_0} = \Lambda_0^2$, the lightest particle is the $(1,1)$
dyon or the $W$-boson
\footnote{For $a_D \simeq 0$, they have practically the same mass,
$\sqrt{2} |a|$.}.  The Seiberg-Witten solution gives, near the
monopole singularity, $|a| \sim \Lambda_0$. Then, in this case, we
have the Wilsonian UV cut-off $\mu = \Lambda_0$.  In general, for an
arbitrary vacuum located at $u$, the order of the cut-off will be
given by $\mu \sim |u - u_i|/\Lambda$, where $u_i$ is the closest
singularity to $u$ (and whose associated BPS state has not been
included in the effective Lagrangian). Notice that in this
``Seiberg-Witten scheme", it is the vacuum who chooses the value of
the Wilsonian UV cut-off.

When supersymmetry is broken down to ${\cal N}=0$, a non-trivial 
effective potential is generated in the effective Lagrangian. 
Its non-supersymmetric terms have the structure
\be
{1 \over \mu^{d_i -4}} {\cal O}_i(\phi) \left({F_0 \over \mu}  
\right)^n,
\label{termsVeff}
\ee
where ${\cal O}_i(\phi)$ is a gauge invariant local operator of 
dimension $d_i$. The Seiberg-Witten solution only gives terms with $n  
\leq 2$.
Higher order terms in $F_0$ come from the non-holomorphic part 
of the effective Lagrangian. We do not include them in our analysis,  
and  from (\ref{termsVeff}) 
we see that our approximation will be valid when $F_0 < \mu \sim |u -  
u_i|/\Lambda$. For these  
values of the supersymmetry breaking scale we can neglect the  
non-holomorphic
terms in  (\ref{termsVeff}). For instance, 
in the soft breaking of the pure ${\cal N}=2$ theory \cite{soft},
with the absolute minimum located very close to the monopole
singularity $u_{\Lambda_0} = \Lambda_0^2$, 
we have that $F_0 < |u_{- \Lambda_0} 
- u_{\Lambda_0}|/\Lambda_0 \sim \Lambda_0$.
In the case of the softly broken, mass-deformed ${\cal N}=4$ 
theory, the distance between the closest Seiberg-Witten 
singularities ${\t u}_i$ is always smaller than $m^2$, and our 
analysis will be done for $F_0<m$. 

The effective potential for the softly broken ${\cal N}=2$ gauge  
theories has the  structure \cite{soft, amz}
\be
V_{\rm eff} = V_{\rm eff}^{(VM)} + V_{\rm eff}^{(HM)} \,.
\label{Veff}
\ee 

The first term is the contribution to the vacuum energy coming 
from the scalar component $a$ of the $U(1)$ vector multiplet, 
and is invariant under macroscopic duality transformations
\cite{soft}. 
It is given by:
\be
V_{\rm eff}^{(VM)} = \Big( {b_{a0}^2 \over b_{aa}} - b_{00} \Big)  
F_0^2 
\label{VeffVM}
\ee
Notice that, because of this invariance, we can choose any duality  
frame to compute the couplings $b_{IJ}$ in  (\ref{VeffVM}). Near the  
$\tilde u_i$ singularity, it is natural to consider 
the duality frame associated to the massless BPS state there, because  
in this frame $b_{a0}^{(i)}$ and $b_{00}^{(i)}$ are smooth.
The positive term $b_{a0}^2/b_{aa}$ increases the vacuum energy, but  
if the coupling $b_{a0}^{(i)}$ is different from zero on its  
associated local singularity $\tilde u_i$, then 
the function (\ref{VeffVM}) presents a local minimum with a cusp
at the point $\tilde u_i$, since $b_{aa}^{(i)} \sim {\rm ln}|\tilde u  
-\tilde  u_i| 
\rightarrow \infty$ for $\tilde u \rightarrow \tilde u_i$ 
(see fig. 3).

The second term in the effective potential is
\be
V_{\rm eff}^{(HM)} = - \sum_{i =1}^3 {2 \over b_{aa}^{(i)}} \rho^4_i
\label{VeffHM}
\ee
where 
\be
\rho^2_i = - b_{aa}^{(i)} |a^{(i)}|^2 +
 {1 \over {\sqrt 2}} |b_{a0}^{(i)}| F_0 \geq 0 \quad \ i=1, 2, 3.
\ee
or zero if this expression is not positive.
This gives the squared VEV of the scalar part of the hypermultiplet  
becoming massless at  
$a^{(i)}({\tilde u}_i) = 0$. If $b_{a0}^{(i)}(\tilde u_i) \not= 0$,  
there is
a condensate $\rho_i \not= 0$ on a region centered at the point
$\tilde u_i$, with radius proportional to $F_0$.  Once the  
hypermultiplet contributions 
are taken into account, the effective potential is smooth on the  
whole $\tilde u$-plane and is 
globally defined (see fig. 4).

\figalign{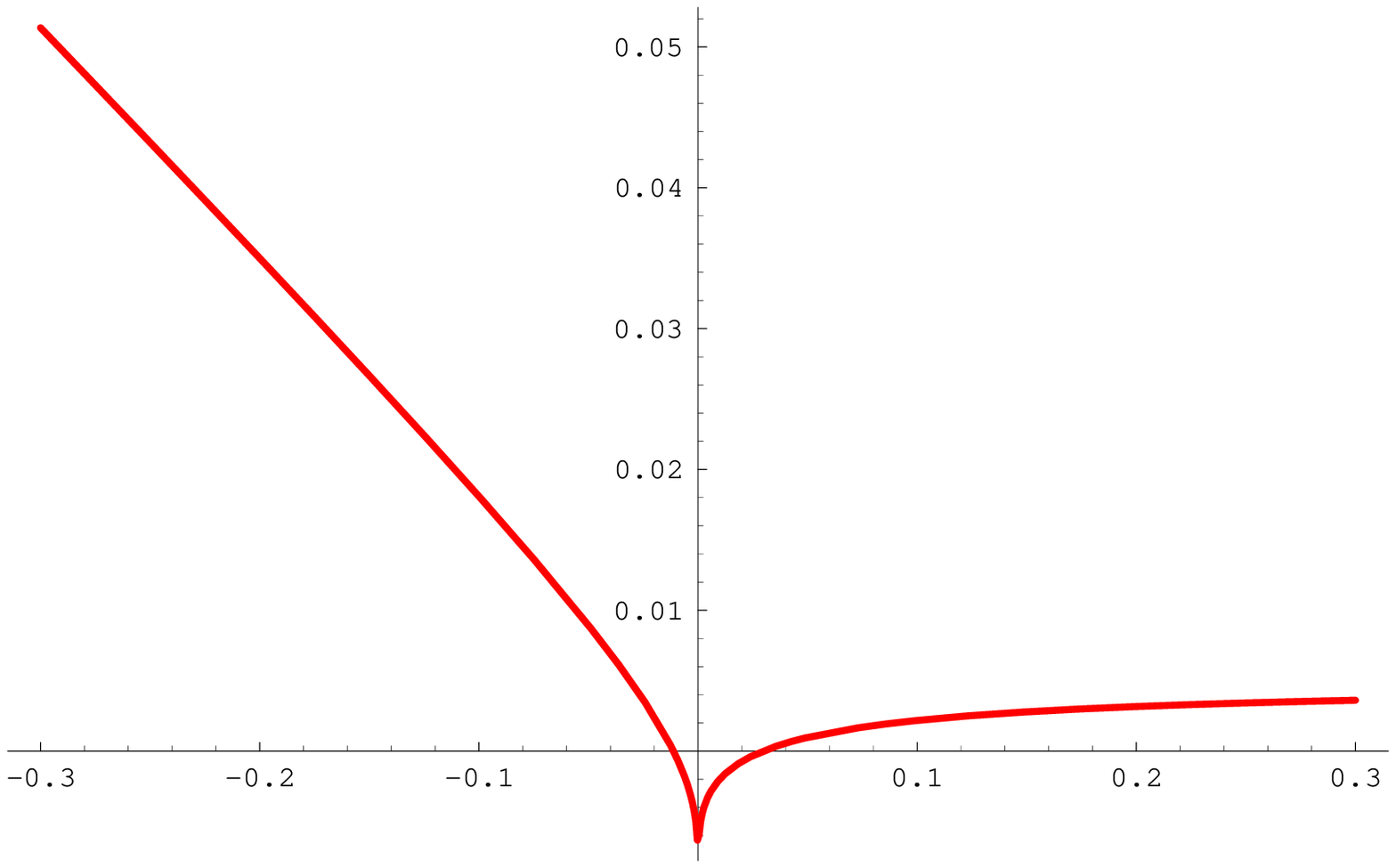}{\B00g}{Plot of $V_{\rm eff}^{(VM)}$
through a path joining the three Seiberg-Witten singularities, for  
$\tau_0=i$.
The only cusp occurs at $\tilde u_3=0$, where $b_{a0}^{(3)}(\tilde  
u_3)\not=0$.}
{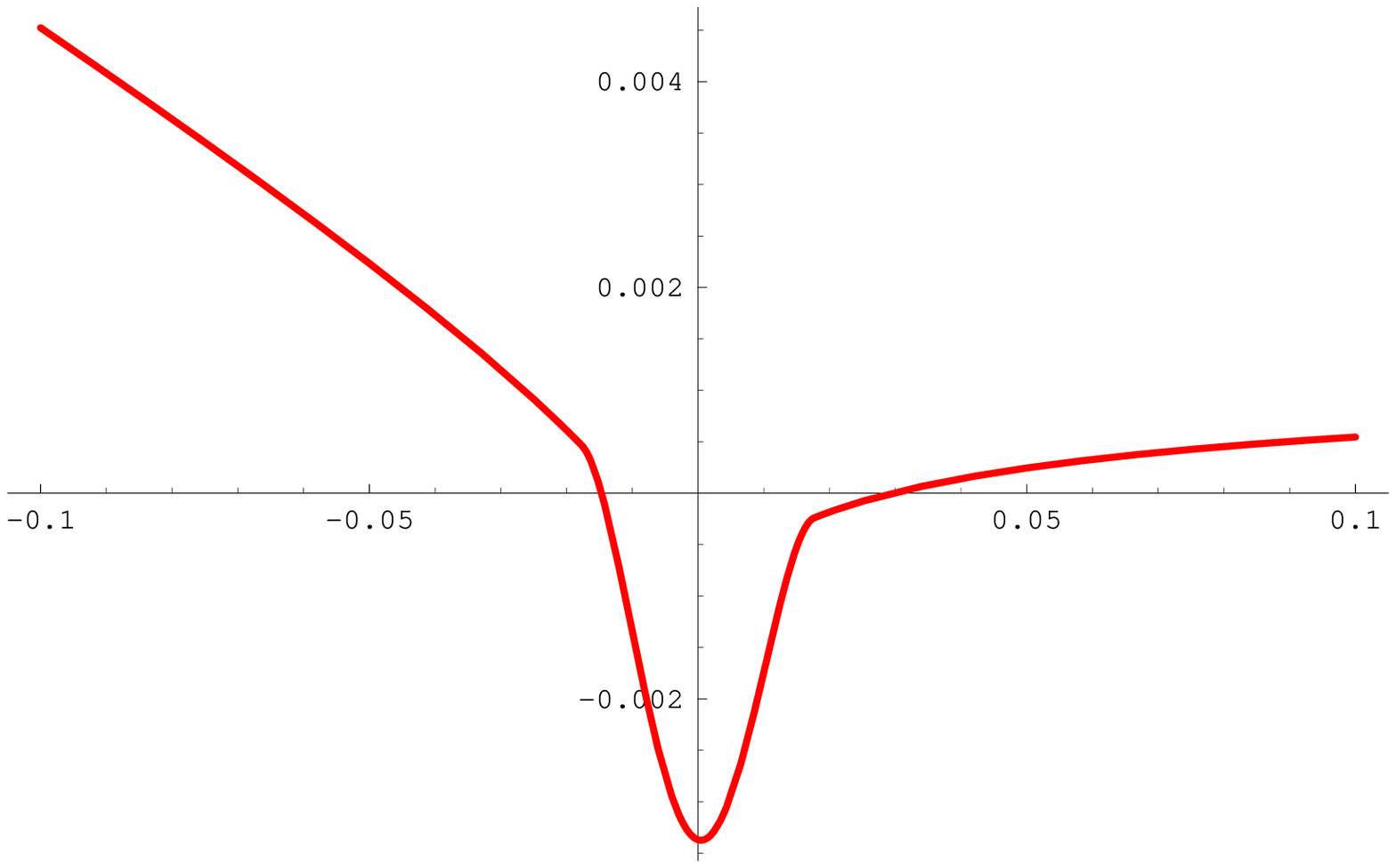}{\Veffg}{Plot of the complete effective potential 
$V_{\rm eff}$ through a path joining the three Seiberg-Witten  
singularities, 
for $\tau_0=i$. The cusp at $\tilde u_3=0$ has been smoothed out by  
the monopole
condensate, and there is an absolute minimum very close to $\tilde  
u_3$.}

The order parameter for the condensation of a light hypermultiplet is  
then given by  
\be
b_{a0}^{(i)}(\tilde u_i) = {1 \over 4 \pi} {\rm Im}\left({\pa {\tilde  
u} \over \pa a^{(i)}}
\right)_{\tilde u=\tilde u_i},
\ee
where the coupling is evaluated at the corresponding singularity 
\footnote{Notice that in the partial breaking to ${\cal N} =1$ of the
Seiberg-Witten solution, the order parameter for condensation is  
$|\pa
{\tilde u}/ \pa a^{(i)} |$.}.  The condensate creates a local minimum
of the effective potential very close to the singularity $\tilde u_i$
(for small values of the supersymmetry breaking parameter), and the
value of the effective potential at that minimum will be given
approximately by $- b_{00}^{(i)}(\tilde u_i)$. When $b_{a0}^{(i)}({\t
u}_i) =0$, although $\rho_i$ can be different from zero near $\tilde
u_i$, the fact that $V_{\rm eff}^{(VM)}$ has no cusp forbids the
formation of a local minimum close to $\tilde u_i$. When there are
several condensate order parameters different from zero on their
associated singularities, we must use numerical information from the
Seiberg-Witten solution to know which singularities give the absolute
minima of the effective potential.

If all the order parameters for condensation turn out to be zero,  
then
there are no local minima near any singularity. In this case, we have
a runaway vacua pehenomenon. To see this, one has to consider the
behavior of the effective potential at infinity, {\it i.e.}, for
$m^2/\tilde u \rightarrow 0$. In this region we are in the
semiclassical regime, and for the self-dual theories there are not
logarithmic (perturbative) corrections in this region of the moduli
space. A straightforward computation gives
\be 
V_{\rm eff} \quad
\rightarrow \quad 8 \pi \alpha_0 |{\rm Im}\sqrt{\tilde u}|F_0^2 \,,
\label{Veffinfty}
\ee 
where $\alpha_0=g_0^2/4 \pi = ({\rm Im}\tau_0)^{-1}$ is the
microscopic gauge coupling.  Hence the effective potential goes to
infinity, except along the direction $\tilde u \rightarrow \infty^+$,
where it goes to zero. Notice that, even when $b_{a0}^{(i)}(\tilde
u_i) \not=0$, the local minima should have a negative vacuum energy  
in
order to give the true minima of the potential. Otherwise, we will
again have runaway vacua.

The expression (\ref{VeffHM}) is just the sum of the contributions of
the different condensates $\rho_i$ to the effective potential, as it
is obtained from an effective Lagrangian that only takes into account
mutually local degrees of freedom at each point $\tilde u$.  But for
values of the supersymmetry breaking parameter big enough, some of the
condensates $\rho_i$ can overlap, and since (\ref{VeffHM}) decreases
the vacuum energy, there is the posibility that a new minimum is
created in the overlapping region.  This could be an indication of a
first order phase transition to an oblique confinement mode, with the
new minima associated to the condensation of a bound state created by
the mutually nonlocal hypermultiplets \cite{soft, amz}.  Actually, in
the mass-deformed ${\cal N}=4$ super Yang-Mills theory, due to the
fact that all the microscopic fields are in the adjoint representation
of the gauge group, the new minimum would be necesarily associated to
an oblique confinement phenomenon \cite{Thooft}.

Even if we allow such overlappings, the region where a condensate,  
call it $\rho_1$, is different from zero,
should not attain any other singularity associated to another state, 
$\tilde u_2$, because 
in this case the effective potential presents a cusp: 
\be
{\pa V_{\rm eff} \over \pa \tilde u} = -{ 8 \over b_{aa}^{(1)}} 
\left({\pa \rho_1^4 \over \pa \tilde u}\right) + \cdots 
\quad \rightarrow \infty
\end{equation}
when $\tilde u \rightarrow \tilde u_2$ (as $b_{aa}^{(1)} \rightarrow  
0$).
This implies that the supersymmetry breaking scale should be smaller  
than
the mass of the BPS associated to $\tilde u_1$ at the point $\tilde  
u_2$, {\it i.e.} 
$F_0 < |\tilde u_1 - \tilde u_2|$. This is the same bound we found 
to neglect higher order corrections in the effective 
Lagrangian, and both things are obviously related. 
To be able to cross this bound, we should include
the mutually nonlocal degrees of freedom asssociated to $\tilde u_1$ 
and $\tilde u_2$ {\it simultaneously} in the effective Lagrangian.

\subsection{Vacuum structure for $\theta_0 = 0$.}
\indent

When the microscopic theta angle is zero, the only order parameter for
condensation which is different from zero is the one associated to the
dyon, $b_{a0}^{(3)}(\tilde u_3)$. In fig. 5 (left), we plot the
absolute value of $b_{a0}^{(3)}(\tilde u_3)$ as a function of the
microscopic coupling $\alpha_0 = g_0^2/(4 \pi) = 1/({\rm Im} \tau_0)$.
The range where it is different from zero is $1/8 {\
\lower-1.2pt\vbox{\hbox{\rlap{$<$}\lower5pt\vbox{\hbox{$\sim$}}}}\ }
\alpha_0 {\
\lower-1.2pt\vbox{\hbox{\rlap{$<$}\lower5pt\vbox{\hbox{$\sim$}}}}\ }
8$.  This means that, for these values of the microscopic gauge
coupling, there is a local minimum of the effective potential very
close to the point $\tilde u_3$.
To be the absolute minimum, it should give a negative vacuum
energy. In fig. 5 (right) we show the value of the effective
potential at $\tilde u_3$, for different values of the microscopic
coupling $\alpha_0$. We then see that there is an absolute minimum
near the point $\tilde u_3$ for $1/3 {\
\lower-1.2pt\vbox{\hbox{\rlap{$<$}\lower5pt\vbox{\hbox{$\sim$}}}}\ }
\alpha_0 {\
\lower-1.2pt\vbox{\hbox{\rlap{$<$}\lower5pt\vbox{\hbox{$\sim$}}}}\ }
3$.  For these values of the microscopic coupling, supersymmetry
breaking selects a unique minimum where the $(1,1)$ BPS state
condenses. The electric charge of this BPS state is actually zero
because of the Witten effect. We then have monopole condensation and
the electric degrees of freedom are confined. The string tension is of
the order of $\sigma \sim \sqrt{|b_{a0}^{(3)}(\tilde u_3)|F_0} \sim
\sqrt{m F_0}$.  There is a mass gap of the same order of magnitude,
and a splitting in the masses of the component fields of the ${\cal
N}=2$ multiplets of the order of $F_0$.  Out of the confinement
region, $1/3 {\
\lower-1.2pt\vbox{\hbox{\rlap{$<$}\lower5pt\vbox{\hbox{$\sim$}}}}\ }
\alpha_0 {\
\lower-1.2pt\vbox{\hbox{\rlap{$<$}\lower5pt\vbox{\hbox{$\sim$}}}}\ }
3$, the theory presents runaway vacua along the positive real axis of
the $\tilde u$-plane.

\begin{figure}
\centerline{
\hbox{\epsfxsize=6cm\epsfbox{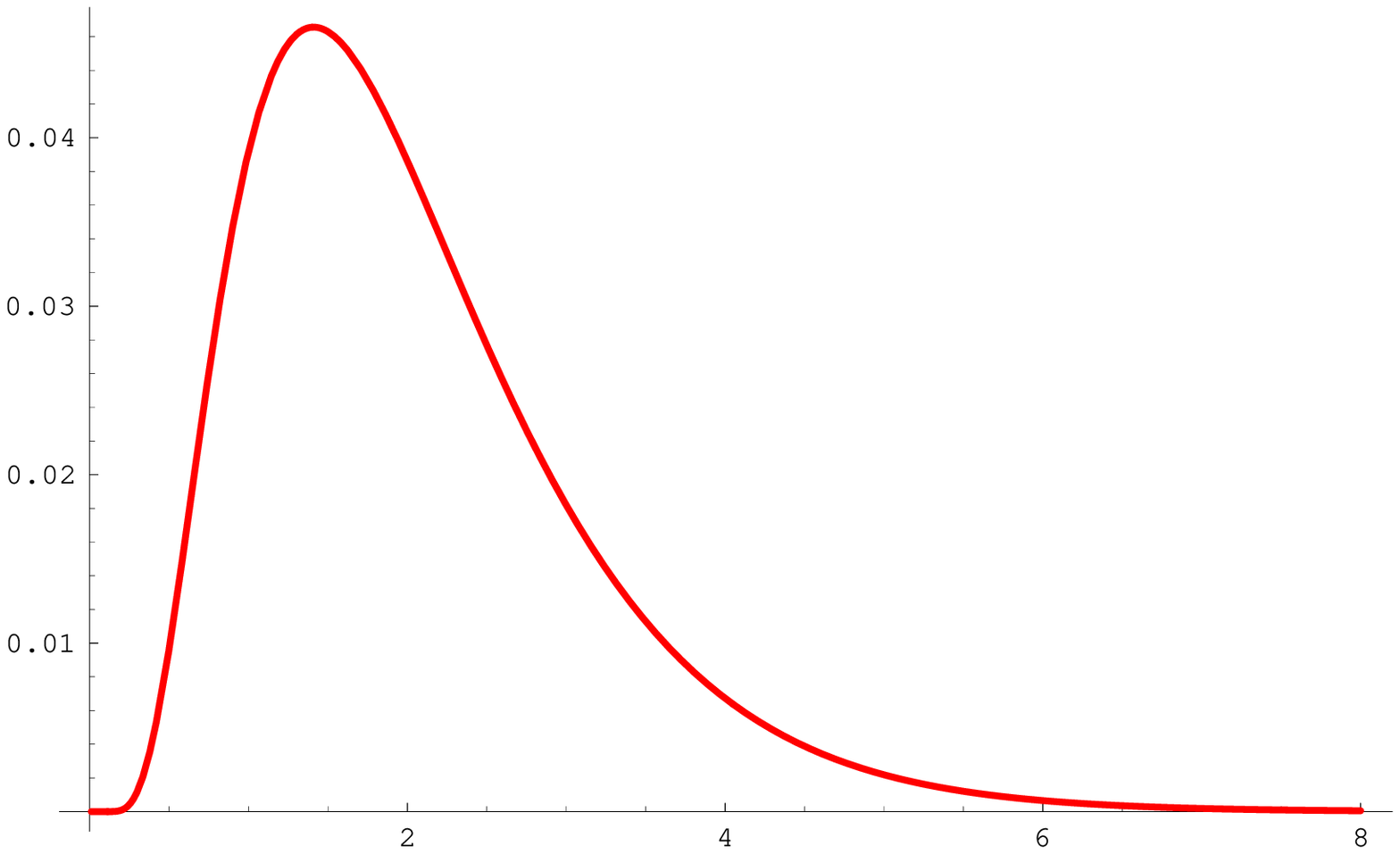}}\qquad
\hbox{\epsfxsize=6cm\epsfbox{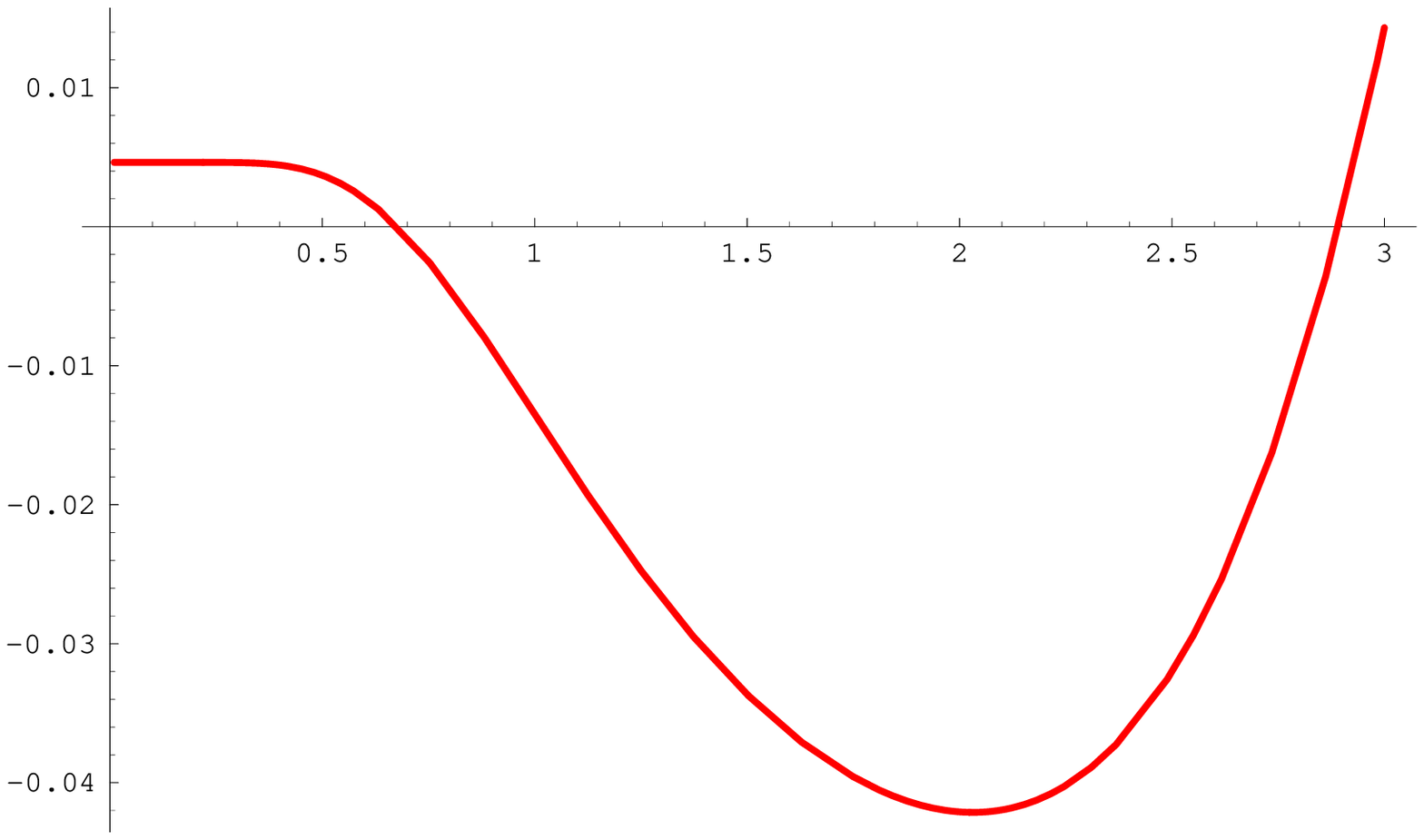}}\qquad}
\caption[]{
The order parameter for monopole condensation, $|b_{a0}^{(3)}(\tilde  
u_3) |$ (left) 
and the effective potential $V_{\rm eff} (\tilde u_3)$ (right), for  
$\theta_0=0$, as a function of the microscopic coupling $\alpha_0$.}
\figlabel\ug
\end{figure}

In the previous discussion, we have chosen a supersymmetry breaking
scale smaller than $m$.  In fact, for a microscopic coupling $\alpha_0
< 1/2$ or $\alpha_0 > 2$, there are always two singularities on the
$\tilde u$-plane which are very close to each other, and this
decreases the maximum allowed value of $F_0$.  At very short
distances, of the order of $1/m$, the description provided by the
Seiberg-Witten effective Lagrangian is no longer valid.

For $\theta_0 =0$, the microscopic $S$-duality maps the theory with  
microscopic coupling 
$\alpha_0 = g_0^2/(4 \pi) = 1/ {\rm Im} \tau_0$ at the 
vacuum ${\tilde u}$ , to the theory 
with inverse microscopic coupling $1/\alpha_0$ at the vacuum $ -  
\alpha_0^2 {\tilde u}$. Since 
${\tilde u}_3 (-1/\tau_0) = \tau_0^2 {\tilde u}_3 (\tau_0)$, we see  
that
the position of the singularity $\tilde u_3$ is invariant (up to the  
modular factor) under $S$-duality.
We also have that
\be
b_{a0}^{(3)} \left(-\alpha^2_0 \tilde u_3; {1 \over \alpha_0} \right)  
= -{ 1  
\over \alpha_0^2} 
b_{a0}^{(3)}(\tilde u_3; \alpha_0). 
\ee
Generically,  the effective potential 
is not left invariant
under $S$-duality; but for $\theta_0=0$, $V_{\rm eff}(\tilde  
u_3;1/\alpha_0)=
(1/\alpha_0^4) V_{\rm eff}(\tilde u_3;\alpha_0)$. The consequence of  
the microscopic $S$-duality symmetry, in the
softly broken theory, is that the physics in the confinement 
phase is almost the same for the couplings $\alpha_0$ and  
$1/\alpha_0$,
up to the scaling factor $\alpha_0$: the softly broken theory  
preserves in some 
way the duality symmetry. This residual symmetry can be seen in  
figure 5: for  
the range of values of the coupling constant where the vacuum is  
stable,   
there are always two values of the coupling constant which give 
the same monopole condensate (and hence the same string tension). 

When the theory is softly broken down to ${\cal N}=1$ \cite{donwit},
all the possible massive vacua are realized for any value of the
coupling constant, and they are permuted by $Sl(2, {\bf Z})$. But, as
we have seen, if all the supersymmetries are softly broken down to
${\cal N}=0$, the vacuum degeneracy is lifted in such a way that the
theory is locked in a confining phase and the only stable vacua occur
for a gauge coupling of order one: the duality symmetry in the phase
structure is lost.

\subsection{Vacuum structure for $\theta_0 \not= 0$.}
\indent

\begin{figure}
\epsfxsize=6.5cm
\centerline{\epsfbox{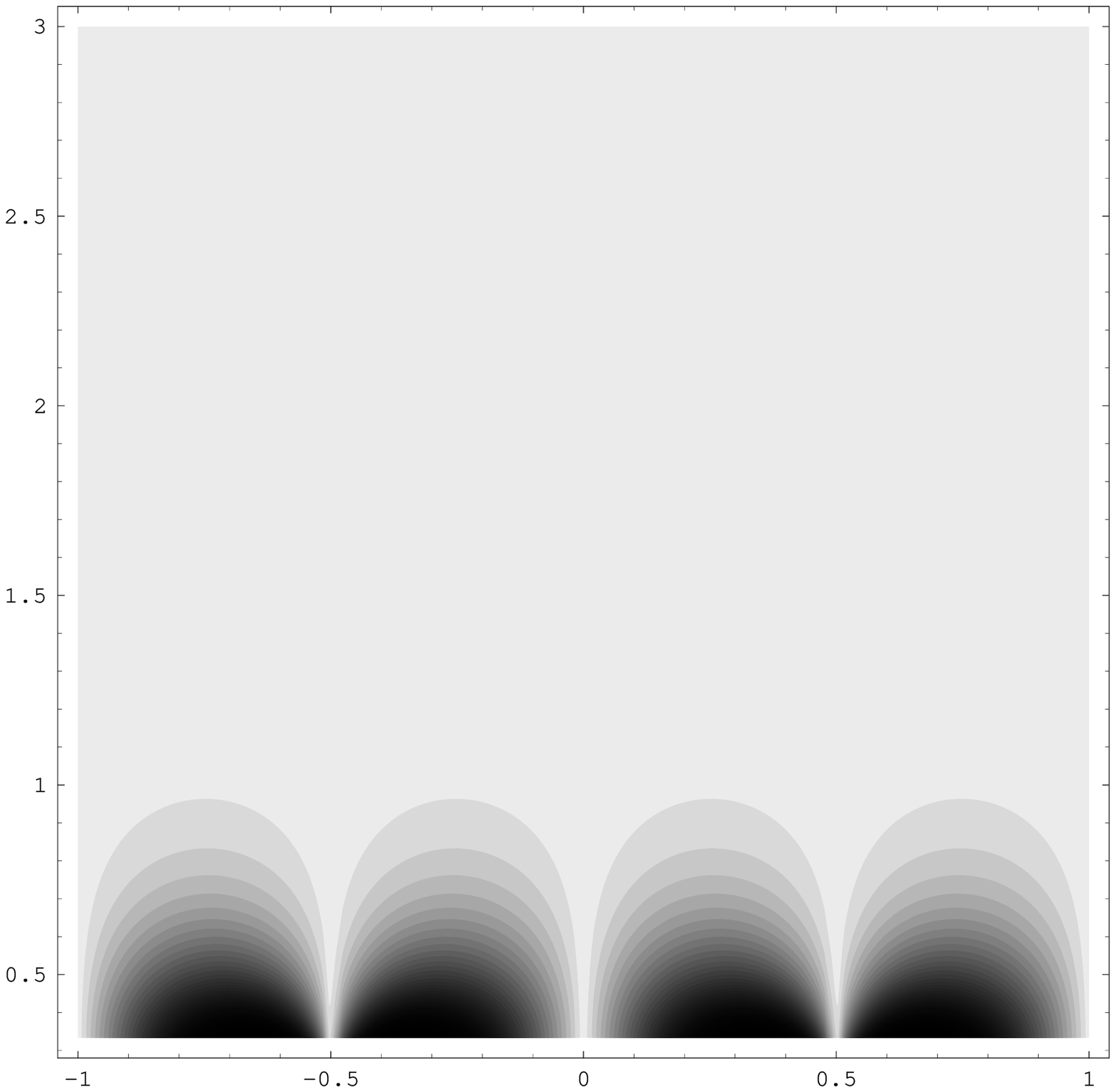}}
\caption[]{Contour plot of the condensate order parameter 
$b_{a0}^{(1)}$, evaluated at the singularity $\tilde u_1$, on the  
complex plane
of the microscopic coupling $\tau_0=x+iy$.}
\figlabel\baq
\end{figure}

In the asymptotically free theories, the $U(1)_R$ anomaly relates the
theory with microscopic theta angle different from zero with the one  
at zero angle but where the $u$-plane is rotated (and the  
rotation angle is proportional to the anomaly). 
In the self-dual theories, there is no $U(1)_R$ 
anomaly, and the theta angle dependence becomes non-trivial.

As a first approach to unravel the theta angle dependence of the
vacuum structure in the softly broken, mass-deformed ${\cal N}=4$
self-dual theory, we obtain the values of the condensate order
parameters.  The contour plots of figures six to eight show the
absolute value of the couplings $b_{a0}^{(i)}$, evaluated at their
respective singularities $\tilde u_i$, for different values of the
microscopic coupling $\tau_0$ in a range $1/3 \leq {\rm Im}\tau_0 \leq
3$ and $-1 \leq {\rm Re}\tau_0 \leq 1$.  Darker zones mean larger
absolute values of the condensate order parameter.

\begin{figure}
\epsfxsize=6.5cm
\centerline{\epsfbox{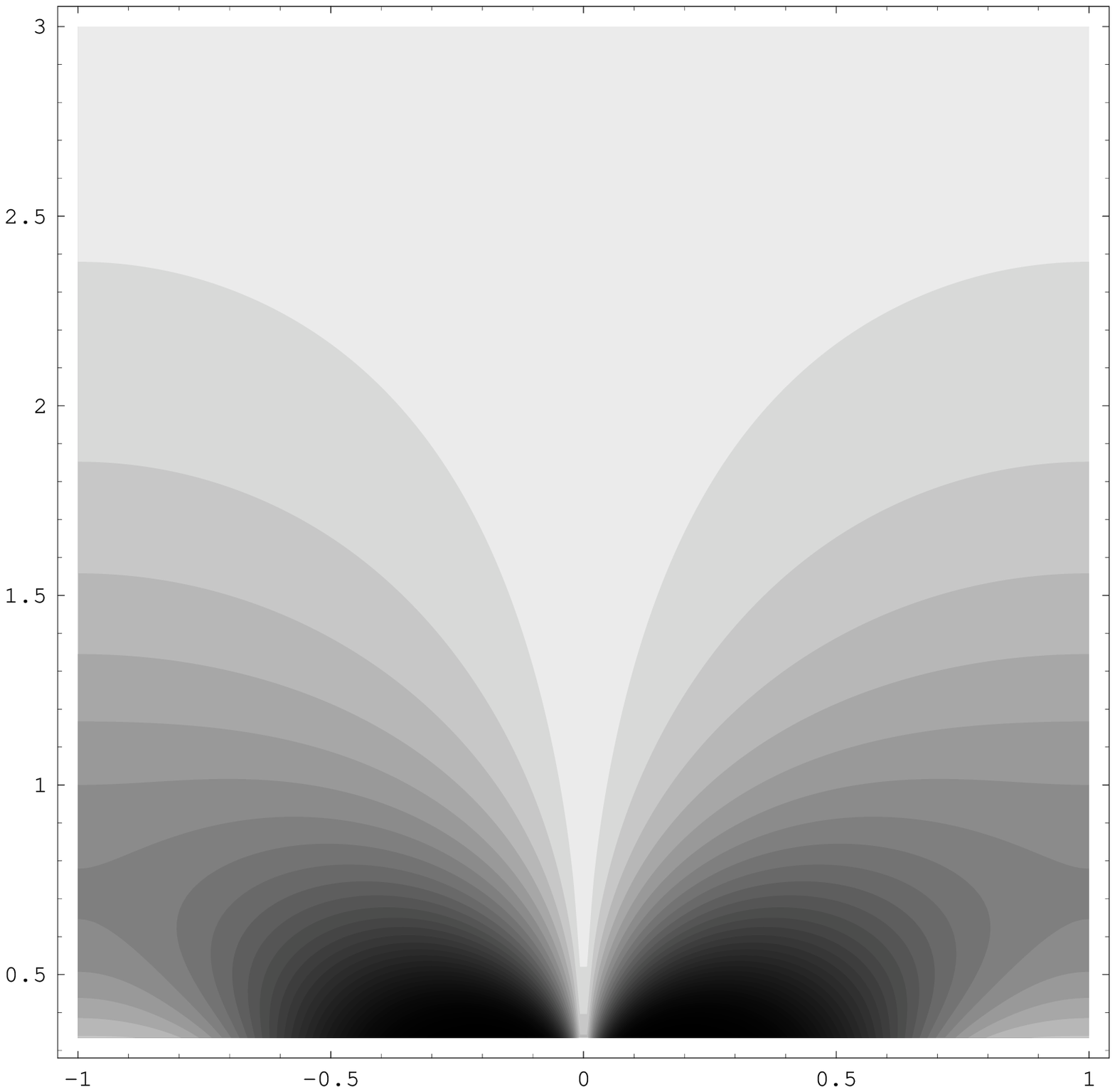}}
\caption[]{Contour plot of the condensate order parameter 
$b_{a0}^{(2)}$, evaluated at the singularity $\tilde u_2$, on the  
complex plane
of the microscopic coupling $\tau_0=x+iy$.}
\figlabel\bam
\end{figure}

\begin{figure}
\epsfxsize=6.5cm
\centerline{\epsfbox{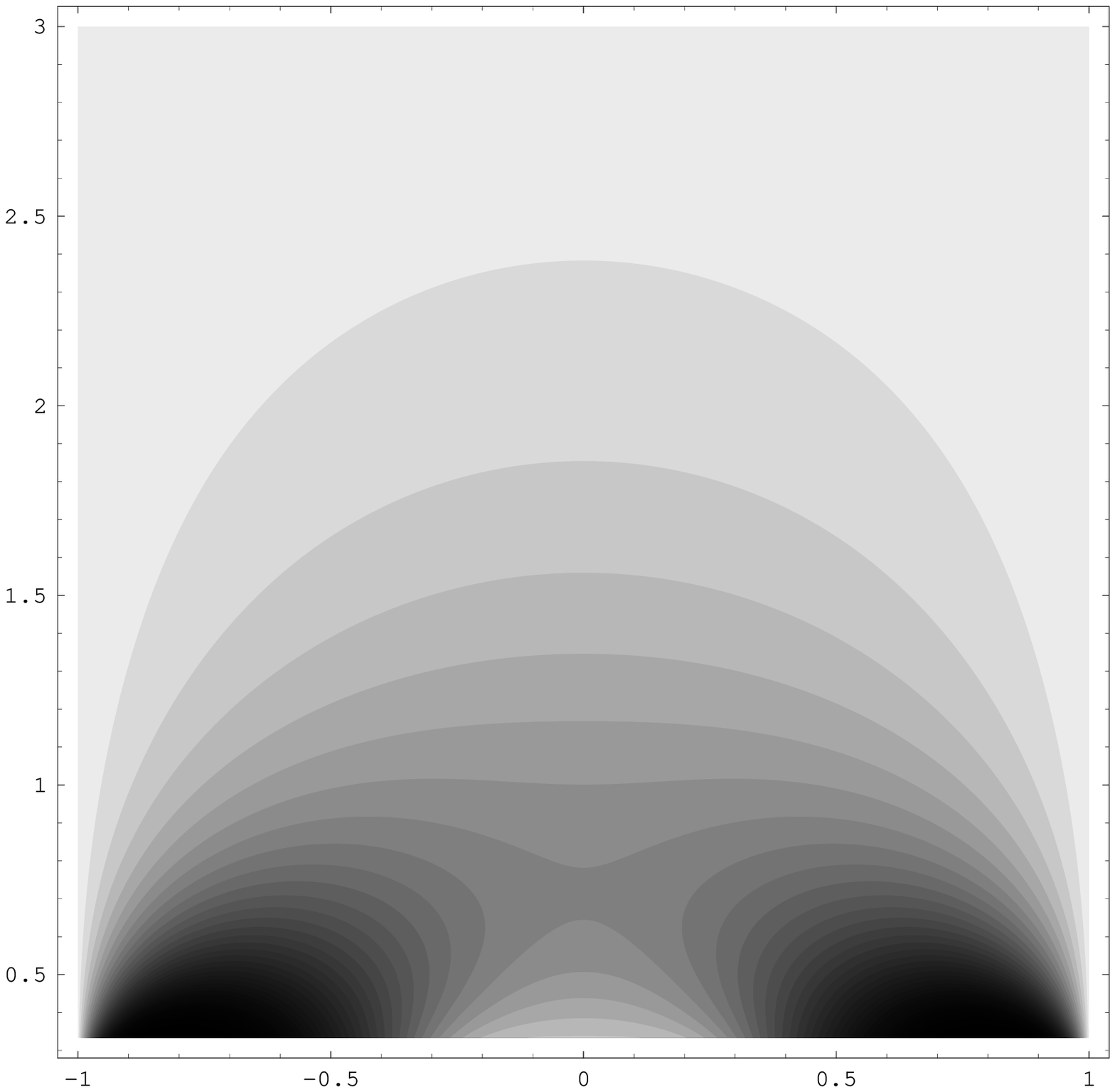}}
\caption[]{Contour plot of the condensate order parameter 
$b_{a0}^{(3)}$, evaluated at the singularity $\tilde u_3$, on the  
complex plane
of the microscopic coupling $\tau_0=x+iy$.}
\figlabel\bad
\end{figure}

At weak coupling (Im$\tau_0 > 3$), all the condensate order parameters
are zero for any value of the microscopic angle. This can be
understood as a remnant of the microscopic $S$-duality and the results
of the previous subsection for zero theta-angle: $S$-duality maps the
theory for general angle $\theta_0$ to the one with angle
$\theta_0^{(S)} = -\alpha_0^2 \theta_0 + {\cal O}(\alpha_0^4)$.  Then,
for small $\alpha_0$, the theory at general $\theta_0$ angle is mapped
to the one at $(1 + \alpha_0^2 \theta_0^2)/\alpha_0$ and $\theta_0
\simeq 0$.  This means that at weak coupling and general $\theta_0$,
there are runaway vacua along $u \rightarrow \infty^+$.  On the other
hand, there is new dynamical information in the strong coupling
region: for finite microscopic theta angle, $S$-duality maps strong
coupling to strong coupling.

Notice that the plot of $b_{a0}^{(2)}(\tilde u_2)$ in \fig\bam\ is
equivalent to the plot of $b_{a0}^{(3)}(\tilde u_3)$ in \fig\bad\ with
the real axis shifted by one unit.  This equivalence is a consequence
of the microscopic $T$-duality symmetry, which interchanges the
singularities $\tilde u_2$ and $\tilde u_3$.  In fact, there is more
information we can extract from $T$-duality, which tells us that
\be
V_{\rm eff}(\tilde u; \tau_0 \pm 1) = V_{\rm eff}(\tilde u; \tau_0),
\ee therefore \be V_{\rm eff}(\tilde u_2; \theta_0=\pi ) = V_{\rm
eff}(\tilde u_3; \theta_0=\pi).  
\ee 

\begin{figure}
\centerline{
\hbox{\epsfxsize=6cm\epsfbox{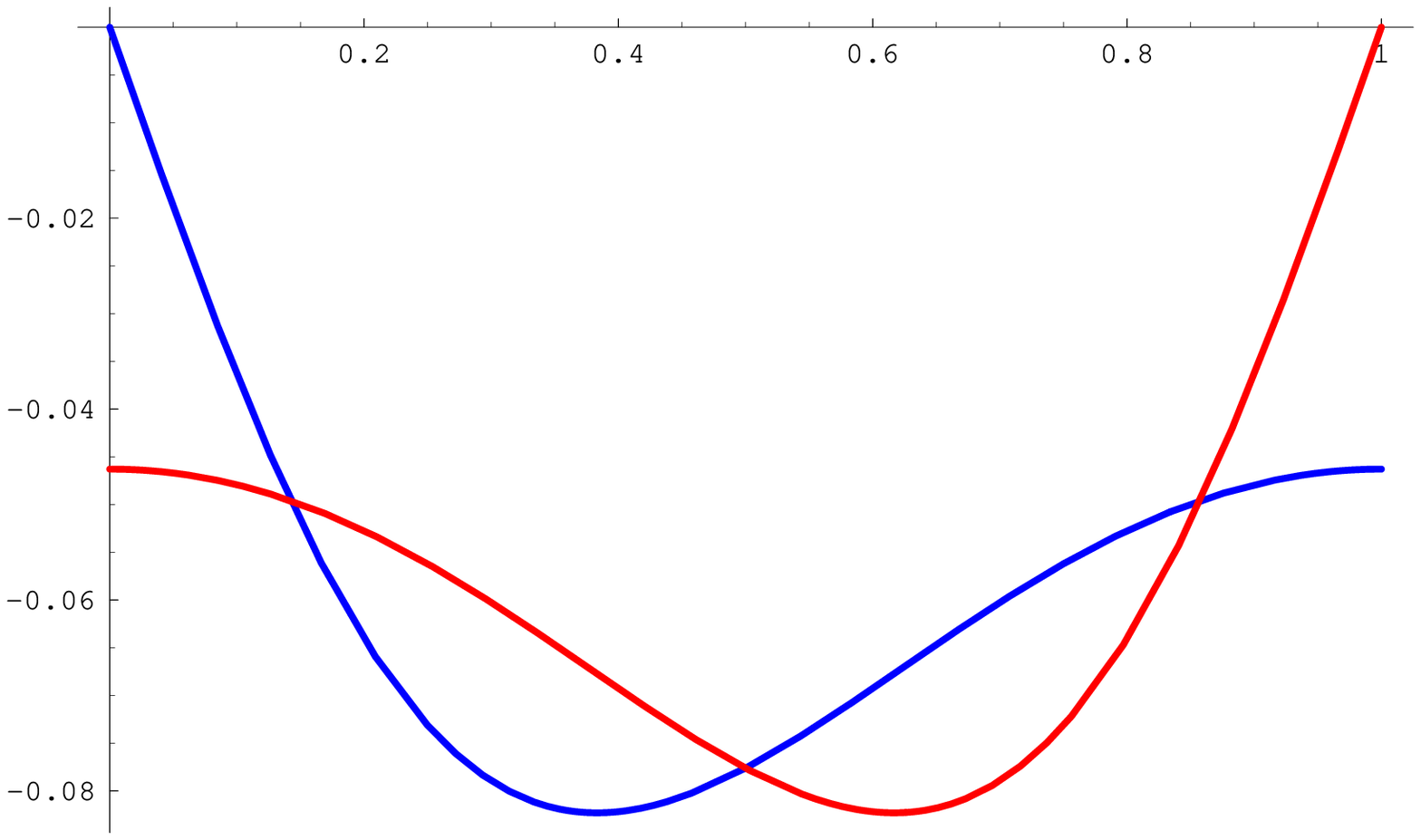}}\qquad
\hbox{\epsfxsize=6cm\epsfbox{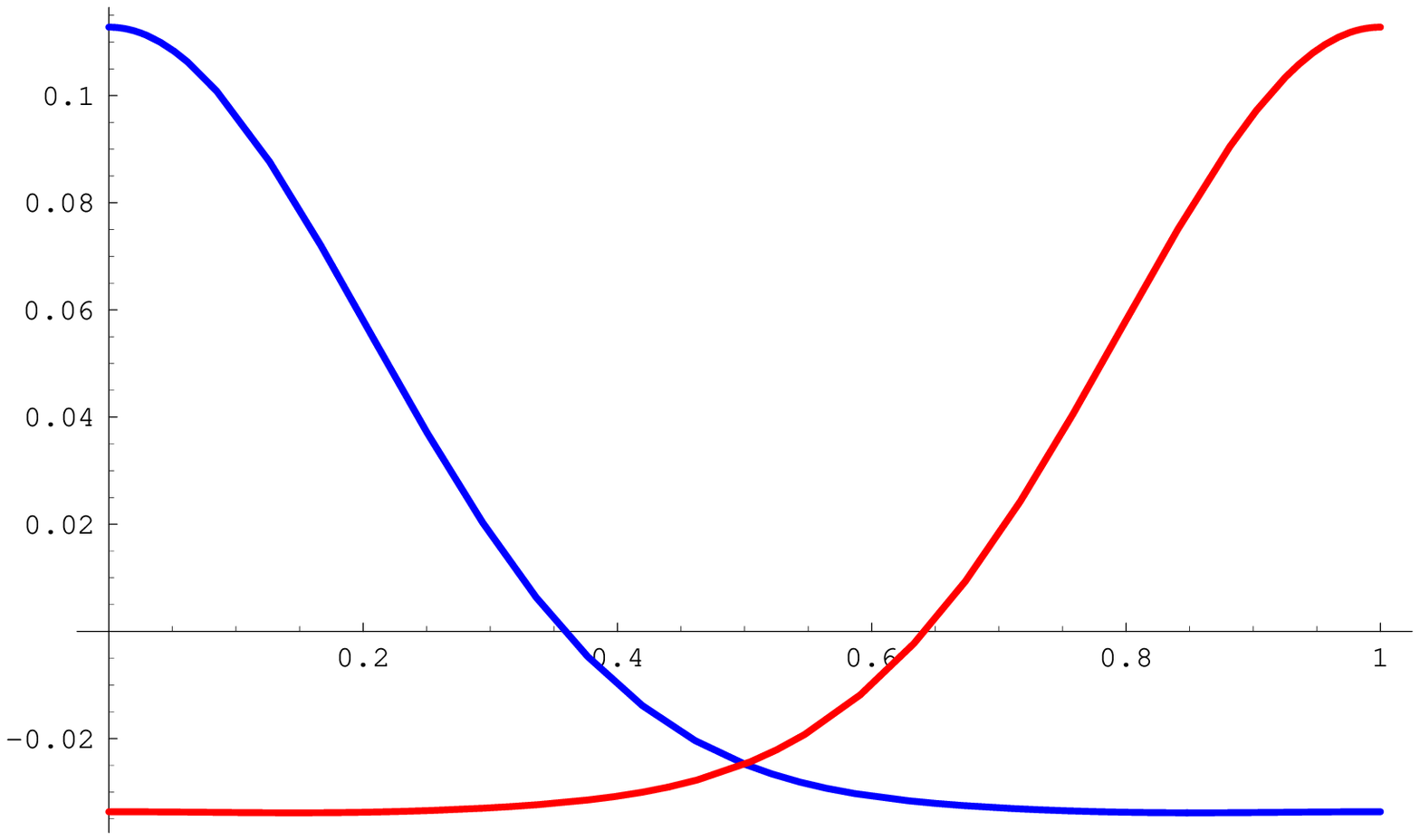}}\qquad}
\caption[]{The condensate order parameters 
$b_{a0}^{(2)}$ and $b_{a0}^{(3)}$ (left) and the effective potential
$V_{\rm eff}$ (right) 
evaluated at their associated singularities $\tilde u_2$ and $\tilde  
u_3$ respectively, 
as a  function of $\theta_0/ 2\pi$, 
for $\alpha_0 =3/2$. The level crossing at $\theta_0=\pi$ corresponds  
to two degenerate minima.}
\figlabel\umud
\end{figure}

We see that, at $\theta_0 =\pi$, if the vacuum energy is smaller than
zero, there must be two equivalent vacua, located at conjugate points
on the $\tilde u$-plane.  Numerically, this happens for $\alpha_0 {\
\lower-1.2pt\vbox{\hbox{\rlap{$>$}\lower5pt\vbox{\hbox{$\sim$}}}}\
}1$.  From the fig. 9, we observe that, when we increase the
microscopic angle $\theta_0 > 0$, it appears a condensate around the
$\tilde u_2$ singularity (giving a new local minimum of the effective
potential), and that the condensate around $\tilde u_3$ decreases.  At
$\theta_0= \pi$, both condensates become equivalent and they give the
same vacuum energy. The physical electric charge of the BPS states
condensing at these two equivalent vacua is exactly zero, but the fact
that the CP transformation interchanges the two complex conjugate
minima at the plane ${\t u}$ is a signal that we have spontaneous CP
symmetry breaking.  In fact there is a first-order phase transition at
$\theta_0=\pi$, and for $\pi <\theta_0 < 2\pi$ the absolute minimum is
located near the $\tilde u_2$ singularity.  This is another test that
the softly broken theory preserves the microscopic $T$-duality
symmetry.  When $\theta_0=2 \pi$, the only condensate occurs around
the $\tilde u_2$ singularity, again with zero electric charge, and it
is located at the same point that the $\tilde u_3$ singularity for
$\theta_0=0$, as it is expected from $T$-duality. A similar behavior
has been observed in other softly broken models when a bare theta
angle is introduced \cite{hsuangle, konishi}.

When $\alpha_0 =2$, the singularity $\tilde u_1$ passes between
$\tilde u_2$ and $\tilde u_3$. As a result, the monodromy matrices of
$\tilde u_1$ and $\tilde u_2$ are conjugated
\footnote{The monodromy matrix of $\tilde u_3$ does not change since
we are working with a monodromy base point ${\t u}_P$ with ${\rm Im}
\tilde u_P > 0$.}. For $\alpha_0>2$ and $\theta_0 \not= n\pi$, the
condensate order parameter $b_{a0}^{(1)}(u_1)$ is different from zero
(see \fig\baq) and there is a new local minimum near ${\t u}_1$. But
numerically it never becomes the absolute minimum of the effective
potential, and the physical vacuum remains located near $\tilde u_2$
or $\tilde u_3$ (or it is a degenerate one for $\theta_0=\pi$).

Although in the deep strong coupling region ($\alpha_0 \gg 1$) the
singularities $\tilde u_2$ and $\tilde u_3$ approach to each other, we
have not found any oblique confinement phase at $\theta_0 = \pi$ (for
the allowed values of the supersymmetry breaking parameter). The same
negative answer was found in other softly broken ${\cal N}=2$ theories
with a bare theta angle \cite{hsuangle}.

\vspace{1.5cm}


{\large\bf Acknowledgements}
\bigskip

We acknowledge L. \'Alvarez-Gaum\'e, J.M.F. Labastida and G. Moore for
a critical reading of the manuscript.  M.M. would like to acknowledge
G. Moore for many useful discussions and remarks, and F. Ferrari for
useful correspondence.  The work of M.M. is supported by DOE grant
DE-FG02-92ER4074. The work of F.Z. is supported by a fellowship from
Ministerio de Educaci\'on y Ciencia.

\newpage

\end{document}